\documentclass[aip,rsi,amsmath,amssymb,preprint]{revtex4-2}

\usepackage{graphicx}
\usepackage{dcolumn}
\usepackage{bm}
\usepackage[utf8]{inputenc}
\usepackage[T1]{fontenc}
\usepackage{mathptmx}
\usepackage{etoolbox}
\usepackage{makecell}
\usepackage{amsmath}
\usepackage{soul}
\usepackage{physics}
\usepackage{xcolor}
\usepackage{multirow}
\usepackage[english]{babel}
\usepackage{placeins}

\begin{document}

\preprint{AIP/123-QED}

\title{Analytical framework for interpreting thermoreflectance measurements of nanoline gratings under finite spot illumination}
\author{Abhishek Pathak}
\altaffiliation[]{These authors contributed equally to the article.}
\author{Seungwon Park}
\altaffiliation[]{These authors contributed equally to the article.}
\author{Shubham Deshmukh}
\author{Alan J. H. McGaughey}
\author{Jonathan A. Malen}
\email[Corresponding Author: ]{jonmalen@andrew.cmu.edu}
\affiliation{Department of Mechanical Engineering, Carnegie Mellon University,\\
5000 Forbes Ave, Pittsburgh, 15213, Pennsylvania, USA.
}

\date{\today}

\begin{abstract}
A frequency-domain thermoreflectance (FDTR)-based technique is presented to measure the effective thermal properties of a material stack whose surface is patterned with metallic nanoline gratings that are heated and probed by lasers.
An analytical heat diffusion model is derived for predicting the FDTR phase response that incorporates a finite laser spot radius with a Gaussian intensity profile for the pump and probe lasers, as well as the dimensions of the nanoline grating. 
The model is validated by comparing with measurements of Au nanoline gratings, with widths ranging from 100 to 1000 nm, fabricated on a-SiO$_2$ and $c$-sapphire substrates.
Due to the short phonon mean free paths (MFPs) in these substrates, thermal transport remains in the diffusive regime for the whole range of nanoline widths.
{\color{black}The model and the measurement show excellent agreement across different laser spot radii and nanoline widths, with a bulk thermal conductivity  value of $c$-sapphire and the thermal boundary conductances at the Au-Cr/a-SiO$_2$ and Au-Cr/$c$-sapphire interfaces comparable to our prior measurements and the literature.}
The validated model will serve as a robust baseline for extending this method to measure effective thermal conductivity as a function of the heater size. This approach is particularly relevant for materials where non-diffusive thermal transport regimes enable probing phonon MFP dependent contributions to the thermal conductivity.
\end{abstract}

\maketitle

\section{Introduction}

When the characteristic length scale $L_c$ (e.g., film thickness, grain size, heat source size) of thermal transport is comparable to or smaller than the phonon mean free paths (MFPs) in a material, the thermal conductivity, $k$, is suppressed relative to its bulk value.  A complete description of how thermal conductivity accumulates with phonon MFPs is given by the material-dependent thermal conductivity accumulation function, $k_{accum}$. Time-domain and frequency-domain thermoreflectance (TDTR and FDTR) experiments have been used on thin films to extract $k_{accum}$ using the film thickness as $L_c$.~\cite{negi2023thickness,zhang2025probing} 
Phonon transmission at interfaces and sample-to-sample variations for samples in a thickness progression, however, can complicate the mapping from the measured thermal conductivity $k_{meas}(L_c)$ to $k_{accum}(L_c)$.  To mitigate these challenges, others used the $L_c$ associated with a heat source size to suppress $k_{meas}$ without altering the physical condition of a bulk sample.
Koh \textit{et al.}~\cite{koh2007frequency} and Regner \textit{et al.}~\cite{regner2013broadband} used the thermal penetration depth, which is inversely proportional to the modulation frequency in TDTR and FDTR, as $L_c$ to measure the size-dependent $k_{meas}$ of bulk substrates.
The interpretation of such measurements is sensitive to the treatment of the interface between the metal transducer and the substrate.~\cite{braun2017upper,regner_advances_2015,regner2014analytical,regner2015interpretation}
Minnich \textit{et al.}~\cite{minnich2011thermal} used the laser spot diameter in TDTR as $L_c$ and demonstrated a suppression of thermal conductivity in silicon as the spot diameter decreased from 60 $\mu$m to 15 $\mu$m.
Johnson \textit{et al.}~\cite{johnson2013direct} applied transient grating (TG) spectroscopy, which uses two intersecting laser beams to form a spatially sinusoidal heating profile in a semiconductor membrane, to measure non-diffusive thermal transport in silicon.
The grating period acts as $L_c$, which can be controlled by changing the angle between the intersecting lasers.~\cite{maznev2011onset}
Cuffe \emph{et al.} used TG spectroscopy to reconstruct the phonon MFP distribution in crystalline silicon~\cite{cuffe2015reconstructing} and Kim \emph{et al.} used a similar approach for amorphous silicon.~\cite{kim2021origin}

The TDTR laser spot diameter and the TG grating period are diffraction-limited. It is therefore difficult to reduce these dimensions below 1 $\mu$m using laser light in the visible spectrum,~\cite{maznev2011onset} thus limiting the range of phonon MFPs that can be suppressed. 
To overcome this limitation, nanoline patterns of metallic heaters can be deposited on a substrate to produce heater sizes and grating periods smaller than the diffraction limit.
Using electron-beam lithography, the width of the individual nanolines can be reduced to tens of nanometers, lowering the lower bound of $L_c$.
Using this approach within the TDTR framework, Zeng \textit{et al.}~\cite{zeng2015measuring} measured the $k_{accum}$ of silicon, employing heaters as small as 30 nm to probe a wide range of phonon MFPs.
Hu \textit{et al.}~\cite{hu2015spectral} used a similar range of nanoline heater sizes to measure $k_{accum}$ and reconstruct the phonon MFP distribution for sapphire, GaN, and GaAs.
Metal nanolines have also been used as diffraction gratings to study quasi-ballistic thermal transport in silicon.~\cite{siemens2010quasi,hoogeboom2015new,frazer2019engineering}
However, in order to achieve similar heater sizes as in previous TDTR studies, this approach often requires the use of ultra-violet or x-ray lasers because the periodicity of the nanolines that can be probed is diffraction-limited.

In this work, we adapt the nanoline grating approach for FDTR measurements.
We derive an analytical solution for the heat diffusion equation for a metal nanoline grating that incorporates the effect of the finite spot radii of the laser beams, which was neglected in previous TDTR studies where the spot radius was large compared to the grating period.
We validate our model through measurements on nanoline patterns fabricated on amorphous fused silica (a-SiO$_2$) and sapphire with the $c$-axis perpendicular to the surface ($c$-sapphire).
These substrates are chosen because their energy carriers exhibit small MFPs in comparison to the nanoline widths and grating periods, to ensure that thermal transport remains in the diffusive regime.\cite{yang2009thermal,zhou_effective_2026}

\section{Analytical model development} \label{sec:analytical}

\begin{figure}
    \centering
    \includegraphics[width=0.9\linewidth]{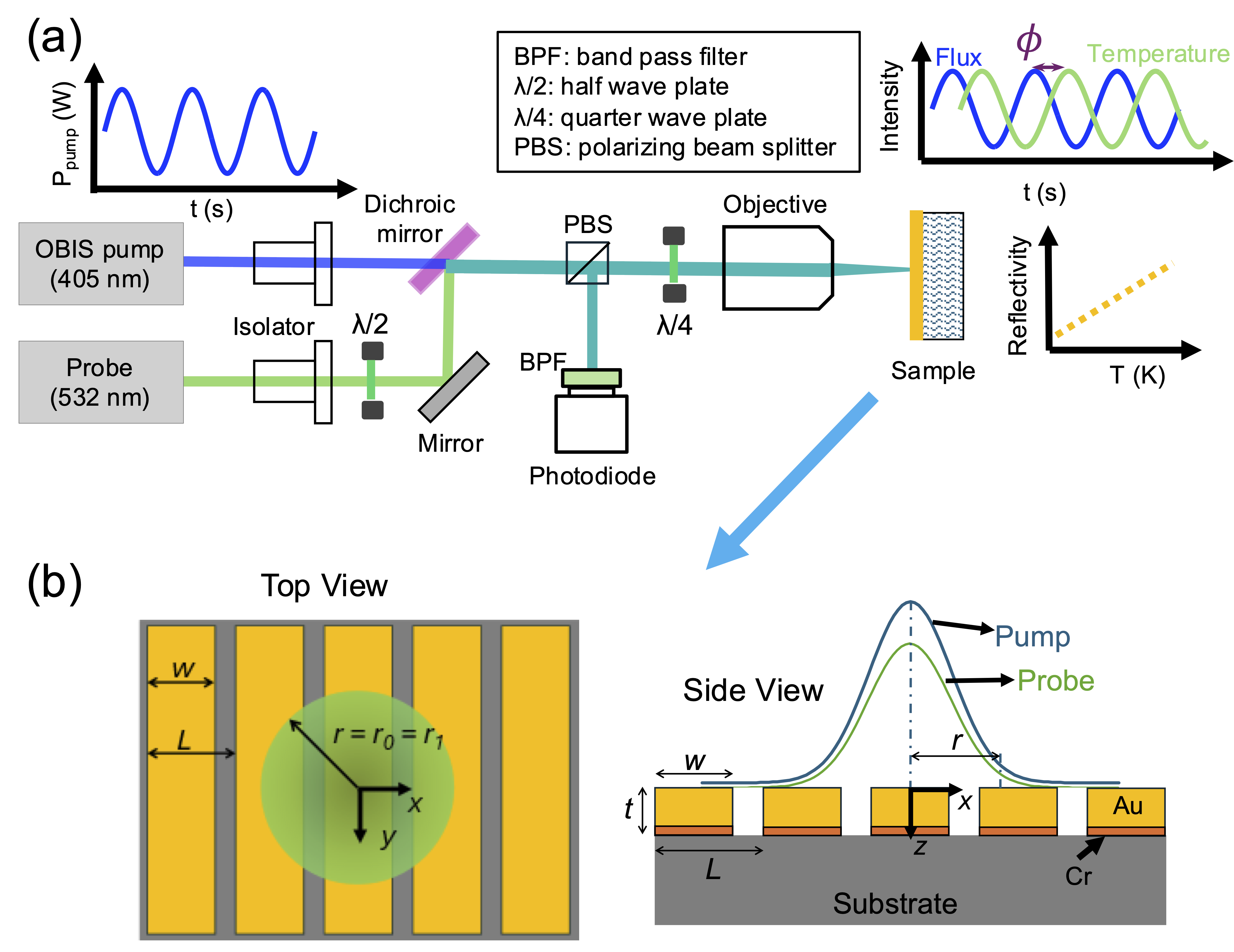}
    \caption{Nanoline array description: \textbf{(a)} Schematic diagram of the frequency-domain thermoreflectance (FDTR) measurement setup with 405 nm pump and 532 nm probe lasers.
\textbf{(b)} Schematic representation of the intensity variation due to the finite spot radius of a Gaussian laser beam over nanolines with a comparable width and separation.}
    \label{fig:description}
\end{figure}

FDTR is a non-contact optical technique in which a continuous-wave pump laser is sinusoidally modulated and focused on the surface of the sample being measured.
The FDTR setup is shown schematically in Fig.~\ref{fig:description}(a).
The top surface of a single or multi-layer sample is coated with a Au film that absorbs the pump laser such that a sinusoidal heat flux is induced at the top surface.
The resulting temperature response on the same surface oscillates sinusoidally with the same frequency ($\omega_{\,0}$) but with a phase lag $\phi(\omega_{\,0})$.
Based on the thermoreflectance of Au, the reflected probe intensity gets modulated with the same $\omega_{\,0}$ and $\phi(\omega_{\,0})$, which can be monitored by a photodiode that is read by a lock-in amplifier.
A thermal transport model for the sample is needed to derive the transfer function $H(\omega_{\,0})$, which relates the thermal and geometric properties of the sample to $\phi(\omega_{\,0})$ as~\cite{schmidt2009frequency} 
\begin{equation}
    \label{eq:phase}
    \phi(\omega_{\,0}) = \tan^{-1}\left\{\frac{\Im[H(\omega_{\,0})]}{\Re[H(\omega_{\,0})]}\right\}.
\end{equation}
This thermal model is then used to fit the desired material and/or geometric properties based on the measured phase lag.
The conventional analytical model used in FDTR~\cite{regner2013broadband} assumes a uniform, continuous metallic film as the transducer.
In this section, we derive an analytical solution of the heat diffusion equation for metal nanoline grating structures.
Our approach builds upon the framework developed by Minnich and co-workers~\cite{Minnich_2011,Minnich_2013} and subsequently adopted by others~\cite{zeng_measuring_2015,hu2015spectral,chen2018quasiballistic} to investigate thermal transport from metal nanoline gratings in TDTR measurements.
The previous measurements approximated the large TDTR laser spot-size with an assumption of uniform spatial heating, as if the spot-size was infinite.
We extend their methodology by incorporating Gaussian intensity variation due to the finite spot radius of both the pump and probe lasers into our analytical model for FDTR. For a large laser spot radius ($r>100\,\mu$m), the phase predicted by our analytical model is identical to that predicted by the infinite spot-size analytical model used in the previous TDTR measurements  (details in Appendix~\ref{sec:appendix_spot_size_effect}).~\cite{hu2015spectral}

We start from the heat diffusion equation of an orthotropic material 
\begin{equation}
\label{eq:governing}
    k_x \frac{\partial^2 T(x,y,z,t)}{\partial x^2} + k_y \frac{\partial^2 T(x,y,z,t)}{\partial y^2} + k_z \frac{\partial^2 T(x,y,z,t)}{\partial z^2} = \rho c \frac{\partial T(x,y,z,t)}{\partial t},
\end{equation}
where $x$ and $y$ are the two in-plane directions and $z$ is the cross-plane direction [Fig.\ref{fig:description}(b)], with $k_x$, $k_y$, and $k_z$ being the directional thermal conductivities. 
$T$, $\rho$, $c$, and $t$ denote temperature, density, specific heat capacity, and time, respectively.
We Fourier transform the $x$, $y$, and $t$ coordinates of Eq.~(\ref{eq:governing}) to obtain an ordinary differential equation in $z$ as 
\begin{multline}
    \label{eq:governing_fourier}
     i^2 \xi_x^2 k_x \tilde{T}(\xi_x,\xi_y,z,\omega) + i^2 \xi_y^2 k_y \tilde{T}(\xi_x,\xi_y,z,\omega)  + k_z \frac{d^2 \tilde{T}(\xi_x,\xi_y,z,\omega)}{d z^2}  = i \omega \rho c \tilde{T}(\xi_x,\xi_y,z,\omega).
\end{multline}
Here, $i^2 = -1$; $\tilde{[\,]}$ denotes the Fourier transform; and $\xi_x$, $\xi_y$, and $\omega$ are the reciprocal space variables corresponding to $x$, $y$, and $t$.
Eq.(\ref{eq:governing_fourier}) can be rearranged as 

\begin{equation}
    \label{eq:governing_compact}
    \frac{d^2 \tilde{T}(\xi_x,\xi_y,z,\omega)}{d z^2} = q^2\tilde{T}(\xi_x,\xi_y,z,\omega),
\end{equation}
where $q=\sqrt{\frac{i \omega \rho c + \xi_x^2 k_x + \xi_y^2 k_y}{k_z}}$. Eq.(\ref{eq:governing_compact}) admits solutions of the form

\begin{equation}
    \label{eq:sol_form}
    \tilde{T}(\xi_x,\xi_y,z,\omega) = \alpha_1 \cosh{qz} + \alpha_2 \sinh{qz},
\end{equation}
where $\alpha_1$ and $\alpha_2$ are constants that are determined based on the boundary conditions.

The bottom surface of a thick sample can be treated as adiabatic. In a stack of materials, the Fourier transformed top-surface [$z=0$ in Fig.~\ref{fig:description}(b)] temperature $ \tilde{T}_t$ and heat flux $\tilde{Q}_t$ can be related to the bottom-surface values ($ \tilde{T}_b$ and $\tilde{Q}_b$) using a transfer matrix approach.~\cite{carslaw1959conduction}
For a material layer $j$ in the stack with thickness $d_j$, the Fourier transformed temperature and heat fluxes at the top and bottom of that layer are related using the quadrupole matrix,~\cite{carslaw1959conduction} $M$ as 
\begin{equation}
    \label{eq:quadrupole}
    \begin{bmatrix}
        \tilde{T}_{b_j}\\
        \tilde{Q}_{b_j}
    \end{bmatrix}
    = M_j \cdot
    \begin{bmatrix}
        \tilde{T}_{t_j}\\
        \tilde{Q}_{t_j}
    \end{bmatrix},
\end{equation}
\text{where}
\begin{equation}
\label{eq:quadrupole_mat}
     M_j = \begin{bmatrix}
        \cosh{(q_jd_j)} & -\frac{1}{k_{zj} q_j}\sinh{(q_jd_j)}\\
        -k_{zj} q_j \sinh{(q_jd_j)} & \cosh{(q_jd_j)}
        \end{bmatrix}.
\end{equation}
For convenience, the explicit dependence of $q$, $\tilde{T}$, $\tilde{Q}$, and $M$ on the variables $\xi_x, \xi_y, \omega, \text{and} \;z$ has been dropped from the notation. 
For a stack with $N$ layers, the relation between temperature and heat flux at the top and bottom of the stack is then
\begin{equation}
    \label{eq:multilayer_matrix}
    \begin{bmatrix}
        \tilde{T}_b\\
        \tilde{Q}_b
    \end{bmatrix}
    =
    \begin{bmatrix}
        \tilde{T}_{b_N}\\
        \tilde{Q}_{b_N}
    \end{bmatrix}
    = M_N M_{N-1} \cdots M_1  \cdot
    \begin{bmatrix}
        \tilde{T}_{t_1}\\
        \tilde{Q}_{t_1}
    \end{bmatrix}
    = 
    \begin{bmatrix}
        A & B \\
        C & D
    \end{bmatrix}
    \cdot
    \begin{bmatrix}
        \tilde{T}_t\\
        \tilde{Q}_t
    \end{bmatrix}.
\end{equation}
Applying an adiabatic boundary condition at the bottom of the stack, which is reasonable for a thick stack, leads to
\begin{equation}
    \label{eq:temp_flux_top}
    \tilde{T}_t = - \frac{D}{C} \tilde{Q}_t.
\end{equation}

$\tilde{Q}_t$ is obtained by a Fourier transform of the heat flux on top of the sample. 
The heat flux on an array of nanolines due to a unit-power Gaussian laser beam, modulated with a temporal frequency $\omega_{\,0}$, is~\cite{cahill2004analysis,Minnich_2011,Minnich_2013}
\begin{equation}
    \label{eq:heat_flux_top}
    \begin{split}
    & Q_t(x,y,t) \, \equiv  \, Q(x,y,z=0,t) = \frac{2}{\pi r_0^2} \exp\left(\frac{-2(x^2+y^2)}{r_0^2}\right) \Bigg[\sum_{n=0}^{\infty} a_n cos\left (\frac{2n\pi x}{L}\right)\Bigg] cos(\omega_{\,0} t),\\ 
    & \text{where } a_n =
    \begin{cases}
        \frac{w}{L} & \text{if } n=0\\
        \frac{2}{n \pi} \sin\left( n \pi \frac{w}{L} \right) & \text{otherwise}.
    \end{cases}      
    \end{split}
\end{equation}
Here, the summation captures the spatial profile of the nanolines, which is infinite in the $y$-direction and periodic in the $x$-direction with periodicity $L$ and nanoline width $w$ [Fig.~\ref{fig:description}(b)].
The location $x=0$ is at the center of a nanoline.
The prefactor describes the Gaussian intensity variation of the pump laser spot with radius $r_0$, which is centered at $(x=0,y=0)$.
After Fourier transforming both spatial coordinates and time, the heat flux is expressed as 
\begin{multline}
    \label{eq:heat_flux_fourier} 
     \tilde{Q}_t(\xi_x,\xi_y,\omega = \omega_{\,0}) \, \equiv  \, \tilde{Q}(\xi_x,\xi_y,z=0,\omega= \omega_{\,0}) \\ \hspace{-125pt} =  \pi  \sum_{n=0}^{\infty} \left\{ \frac{a_n}{2} \exp\left(\frac{-\xi_y^2r_0^2}{8}\right) \right. \\ \left. \times \left[\exp\left(\frac{-(\xi_x - \frac{2n\pi}{L})^2r_0^2}{8}\right) + \exp\left(\frac{-(\xi_x + \frac{2n\pi}{L})^2r_0^2}{8}\right)\right]\right\}.
\end{multline}

Inserting Eq.~(\ref{eq:heat_flux_fourier}) into Eq.~(\ref{eq:temp_flux_top}) gives
\begin{multline}
    \label{eq:temp_fourier}
    \tilde{T}_t(\xi_x,\xi_y,\omega) \, \equiv  \,\tilde{T}(\xi_x,\xi_y,z=0,\omega) \\ \hspace{-115pt} = \pi \delta(\omega - \omega_{\,0})\sum_{n=0}^{\infty} -\frac{D}{C} \left\{ \frac{a_n}{2} \exp\left(\frac{-\xi_y^2r_0^2}{8}\right) \right. \\ \left.\times \left[\exp\left(\frac{-(\xi_x - \frac{2n\pi}{L})^2r_0^2}{8}\right) + \exp\left(\frac{-(\xi_x + \frac{2n\pi}{L})^2r_0^2}{8}\right)\right]\right\}.
\end{multline}
Here $\delta(\omega-\omega_{\,0})$ signifies that for a sinusoidal heat flux with a single frequency component $\omega_{\,0}$, the temperature response recorded at the photodetector will only have the same frequency component.
Taking the inverse Fourier transform of Eq.~(\ref{eq:temp_fourier}) in $x$ and $y$ leads to the real-space temperature field oscillating at the modulation frequency $\omega$,
\begin{equation}
    \label{eq:temp_inverse_transform}
    T_t(x,y,\omega) = \int_{-\infty}^{\infty} \int_{-\infty}^{\infty} \tilde{T}(\xi_x,\xi_y,\omega) \exp(i\xi_x x) \exp(i\xi_y y)d\xi_x d\xi_y.
\end{equation}
The probe laser also has a Gaussian intensity profile with a spot radius $r_1$, centered at $(x=0,y=0)$ at the surface of the sample.
The probe Gaussian intensity profile multiplied by the spatial profile of the nanolines acts as a weighting factor for the thermoreflectance response from the top surface of the sample. The transfer function $H(\omega)$ quantifies how this spatial weighting maps the measured thermoreflectance to the original pump modulation signal.
The corresponding transfer function is
\begin{multline}
    \label{eq:transfer_function_def}
    H(\omega)=\int_{-\infty}^{\infty} \int_{-\infty}^{\infty} T_t(x,y,\omega) \times \frac{2}{\pi r_1^2} \exp\left(\frac{-2(x^2+y^2)}{r_1^2}\right) \sum_{m=0}^{\infty}a_m \cos{\left(\frac{2m\pi x}{L}\right)} dx dy.
\end{multline}
The Fourier coefficients $a_m$ have the same form as the $a_n$ in Eq.~(\ref{eq:heat_flux_top}) as they describe the same nanoline pattern. Substituting Eq.~(\ref{eq:temp_inverse_transform}) into Eq.~(\ref{eq:transfer_function_def}) leads to the a four-dimensional integral expression for $H(\omega)$,
\begin{multline}
    \label{eq:transfer_function_4D}
    H(\omega) =  \pi \delta(\omega - \omega_{\,0})\sum_{n=0}^{\infty} \sum_{m=0}^{\infty} \int_{-\infty}^{\infty} \int_{-\infty}^{\infty}\int_{-\infty}^{\infty} \int_{-\infty}^{\infty} \frac{a_ma_n}{\pi r_1^2}\left( -\frac{D}{C}\right) \\ \times \left[\exp\left(\frac{-(\xi_x-\frac{2n\pi}{L})^2 r_0^2}{8}\right) + \exp\left(\frac{-(\xi_x + \frac{2n\pi}{L})^2 r_0^2}{8}\right)\right] \\ \times \exp \left( i\xi_x x + i \xi_y y -\frac{ \xi_y^2 r_0^2}{8} - \frac{2(x^2+y^2)}{r_1^2}\right)  \times \cos{\left( \frac{2m\pi x}{L}\right)} d\xi_x d\xi_y dx dy.
\end{multline}

As shown in Appendix~\ref{sec:appendix_4d_to_2d}, Eq.~(\ref{eq:transfer_function_4D}) can be converted from a four-dimensional integral to a two-dimensional integral using Hubbard-Stratonovich transformation~\cite{schafer1980disordered} to obtain
\begin{multline}
    \label{eq:main_2D_transfer_function_unequal}
    H(\omega) = \pi \delta(\omega-\omega_{\,0})\sum_{n=0}^{\infty} \sum_{m=0}^{\infty} \int_{-\infty}^{\infty} \int_{-\infty}^{\infty} \frac{a_na_m}{4} \left(-\frac{D}{C}\right) \exp \left[-\frac{ \xi_y^2}{8} (r_0^2 + r_1^2)\right] \\ \times \left\{\exp\left[ -\frac{r_0^2}{8} \left(\xi_x-\frac{2n\pi}{L}\right)^2 -\frac{r_1^2}{8} \left(\xi_x-\frac{2m\pi}{L}\right)^2\right] \right. \\ \left. + \exp\left[ -\frac{r_0^2}{8} \left(\xi_x-\frac{2n\pi}{L}\right)^2 -\frac{r_1^2}{8} \left(\xi_x+\frac{2m\pi}{L}\right)^2\right] \right. \\ \left. + \exp\left[ -\frac{r_0^2}{8} \left(\xi_x+\frac{2n\pi}{L}\right)^2 -\frac{r_1^2}{8} \left(\xi_x-\frac{2m\pi}{L}\right)^2\right] \right. \\ \left. + \exp\left[ -\frac{r_0^2}{8} \left(\xi_x+\frac{2n\pi}{L}\right)^2 -\frac{r_1^2}{8} \left(\xi_x+\frac{2m\pi}{L}\right)^2\right]\right\}  d\xi_x d\xi_y.
\end{multline}
{\color{black}The pump and probe laser beam radii were measured independently using the knife-edge method and were found to agree within 10\%. Therefore, we} assume that the pump and probe lasers have the same spot radius on the surface of the sample, $r_0 = r_1 = r$, which simplifies Eq.~(\ref{eq:main_2D_transfer_function_unequal}) into
\begin{multline}
    \label{eq:2D_transfer_function}
    H(\omega) = \pi \delta(\omega-\omega_{\,0})\sum_{n=0}^{\infty} \sum_{m=0}^{\infty} \int_{-\infty}^{\infty} \int_{-\infty}^{\infty} \frac{a_na_m}{4} \left(-\frac{D}{C}\right) \exp \left(-\frac{ \xi_y^2}{4} r^2\right) \\ \times \left\{\exp\left[ -\frac{r^2}{8} \left(\xi_x-\frac{2n\pi}{L}\right)^2 -\frac{r^2}{8} \left(\xi_x-\frac{2m\pi}{L}\right)^2\right] \right. \\ \left. + \exp\left[ -\frac{r^2}{8} \left(\xi_x-\frac{2n\pi}{L}\right)^2 -\frac{r^2}{8} \left(\xi_x+\frac{2m\pi}{L}\right)^2\right] \right. \\ \left. + \exp\left[ -\frac{r^2}{8} \left(\xi_x+\frac{2n\pi}{L}\right)^2 -\frac{r^2}{8} \left(\xi_x-\frac{2m\pi}{L}\right)^2\right] \right. \\ \left. + \exp\left[ -\frac{r^2}{8} \left(\xi_x+\frac{2n\pi}{L}\right)^2 -\frac{r^2}{8} \left(\xi_x+\frac{2m\pi}{L}\right)^2\right]\right\}  d\xi_x d\xi_y.
\end{multline}

\section{FDTR Measurement}\label{sec:FDTR}
A schematic diagram of the FDTR setup is shown in Fig.~\ref{fig:description}(a).
The pump laser (Coherent OBIS 405 nm LX) emits an intensity-modulated beam, with a frequency range of 100 kHz to 5 MHz chosen for these experiments. 
The pump laser serves as a temporally periodic, spatially Gaussian heat source. 
The pump laser is absorbed by the Au nanolines, inducing a temperature response that exhibits a phase lag relative to heating, which depends on the thermal properties of the substrate. 
Based on the thermoreflectance of Au, the phase lag is monitored by a co-aligned probe laser (Coherent Verdi-V6 532 nm) for each modulation frequency. 
\begin{figure}[!h]
    \centering
    \includegraphics[width=0.9\linewidth]{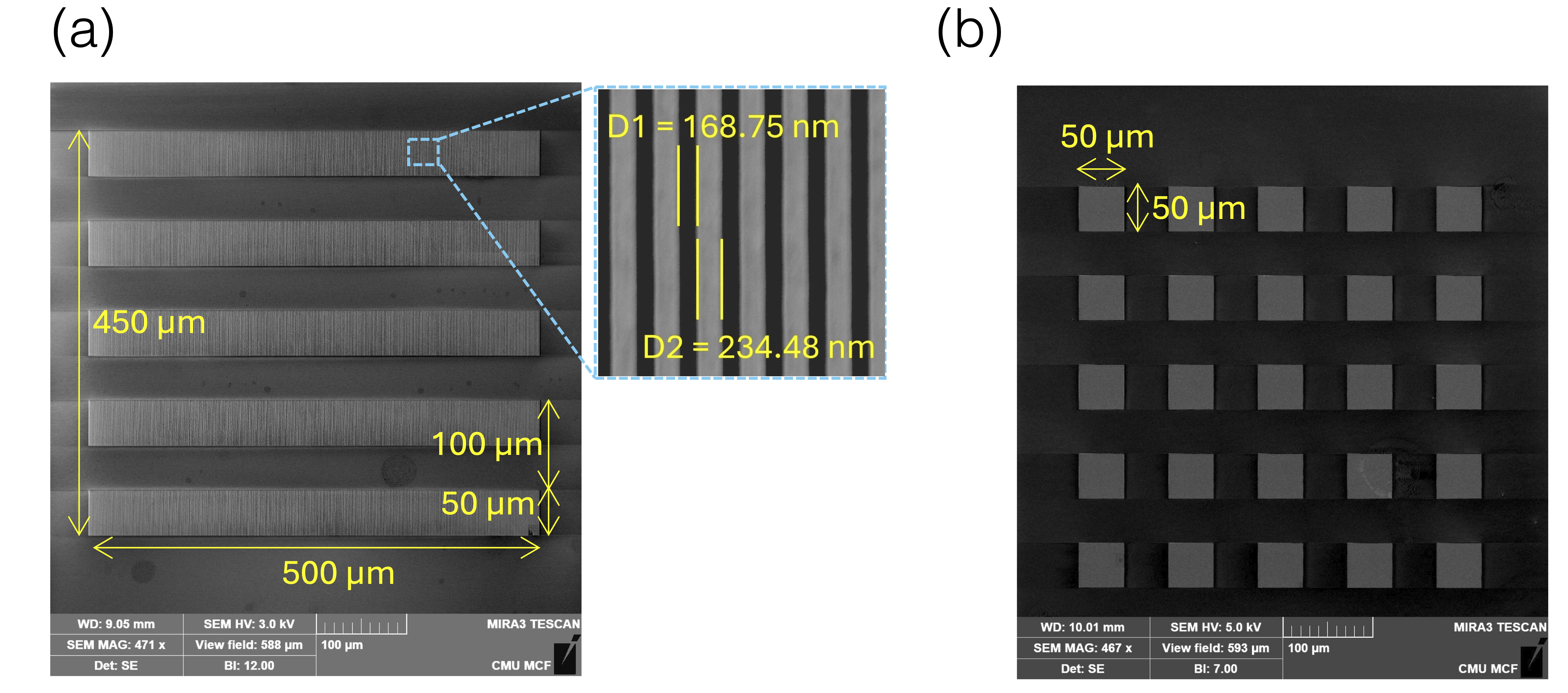}
    \caption{\textbf{(a)} SEM image of a nanoline pattern. D1 represents gap $= L-w$, and D2 represents $w$. \textbf{(b)} SEM image of 5$\times$5 matrix of square patches.}
    \label{fig:nanofab}
\end{figure}
 
\renewcommand{\arraystretch}{1.3}
\setlength{\tabcolsep}{12pt} 
\begin{table}[htbp]
    \caption{Measured nanoline patterns.}
    \begin{center}
        \begin{tabular}{c|cc}
        \hline
    Substrate & $w$ (nm) & gap $= L- w$ (nm) \\
    \hline
    \multirow{3}{*}{a-SiO$_2$} & 172 & 180\\
    & 524 & 170 \\
    & 1024 & 164 \\
    \hline
     \multirow{3}{*}{$c$-sapphire} & 234 & 169\\
    & 548 & 153 \\
    & 1052 & 154 \\
    \hline
        \end{tabular}
        \label{nanoline_list}
    \end{center}
    \label{table:nanoline}
\end{table}
{\color{black}The process by which the nanolines were deposited on the substrates is described in Sec. S1. The nanoline width and spacing were characterized using scanning electron microscopy [Fig. \ref{fig:nanofab}(a)], while thickness was measured using a KLA Tencor P-15 Profilometer, as shown in Sec.~\ref{sec:SI_profilometer} of the Supplementary Information.
The dimensions of all nanoline patterns measured in this study are listed in Table~\ref{nanoline_list}. In addition to the nanolines, a 5$\times$5 matrix of square Au patches, each 50$\times$50 $\mu$m$^2$, was fabricated on the substrate as a control, as shown in Fig.~\ref{fig:nanofab}(b).
The lateral dimensions of these patches are much larger than the laser spot radius ($r<5$ $\mu$m), allowing the application of the standard FDTR analytical solution for interpreting the measurements on them.

On the nanoline patterns, t}he pump and probe beams passing through the gap between the nanolines do not directly affect the thermoreflectance response since both a-SiO$_2$ and $c$-sapphire are transparent to the pump and probe wavelengths.
As a confirmation, no thermoreflectance response was obtained on a $c$-sapphire substrate in the absence of a Au transducer, as shown in Fig.~\ref{fig:SI_thermoreflectance_sapphire}.
We did observe that the phase lag was affected by the portion of the pump laser beam that passed through the nanoline gaps and was reflected off the rear surface of both substrates, causing backside heating.
{\color{black} The effect was observed as a significant departure of the measured phase lag from the prediction of the thermal model.}
To mitigate this issue, we applied a thin layer of glycerin, which has a refractive index ($\approx$ 1.5) similar to that of a-SiO$_2$ and $c$-sapphire, on the rear surface of the substrates. 
This thin glycerin layer effectively eliminated internal reflection by allowing the light that passed through the grating to exit through the back surface of the substrate and scatter diffusively in the glycerin.
{\color{black} Upon application of the glycerin layer, the measured phase lag became consistent with the thermal model.}

The measured phase lag as a function of the modulation frequency was fitted using the analytical model presented in Sec.~\ref{sec:analytical} to extract the unknown thermal properties. 
The in-plane thermal conductivity of the Au nanolines in the direction perpendicular to the nanoline orientation ($k_{x,Au}$) was set to 0 W/m·K to simulate the physical gap between the nanolines.
{\color{black}In Sec.~\ref{sec:nanoline_FEM}, we present results from a 3D finite element study to show that the predicted phase lag is insensitive to this simplification.}
On the a-SiO$_2$ substrate, we extracted the in-plane thermal conductivity of the Au nanolines in the direction parallel to the nanoline orientation ($k_{y,Au}$) and the thermal boundary conductance at the Au-Cr/a-SiO$_2$ interface ($G_{Au-Cr/a-SiO_2}$). 
For the Au nanolines on the $c$-sapphire substrate, we extracted the thermal conductivity of $c$-sapphire {\color{black}($k_{c-sapphire}$)} and the thermal boundary conductance at the Au-Cr/$c$-sapphire interface ($G_{Au-Cr/c-sapphire}$). {\color{black} A single $k_{c-sapphire}$ value was extracted because its in-plane (parallel to $c$-plane) and cross-plane (normal to $c$-plane) thermal conductivities are comparable at room temperature. \cite{tang_anisotropic_2021, rahman_simultaneous_2019}}

Two Mitutoyo objective lenses with different magnifications, 10$\times$ and 20$\times$, were used to explore the effect of the laser spot radius, which is a critical parameter in the analytical model.
We used the square patch arrays on both substrates as reference samples to measure the laser spot radius.
The phase lag was fitted to the standard multilayer solution of FDTR~\cite{regner2013broadband} to extract the laser spot radius and the thermal boundary conductance at the Au-Cr/substrate interface ($G_{Au-Cr/sub}$). Since the analytical FDTR model does not explicitly include the thermal boundary conductance, the interface was modeled as a thin layer with thickness $d=1$\,\AA{} and thermal conductivity $k$, where $G = k/d$. 
Since the nanolines were deposited on the same substrate under the same conditions and within the same fabrication process, $G_{Au-Cr/sub}$ of the square patches should be the same as $G_{Au-Cr/sub}$ for the nanolines.
The laser spot radius was then treated as a fixed parameter for the particular objective used during the nanoline measurements. 
For the nanoline measurements, the sample stage was positioned with the micro-manipulator so that the probe laser was entirely within the nanoline pattern, with the laser center aligned to the center of one of the nanolines.

\section{Results}\label{sec:results}
\subsection{a-SiO$_2$}\label{subsec:a-SiO2results}
We first studied a-SiO$_2$ to experimentally validate the analytical solution derived in Sec. \ref{sec:analytical}. Phonon MFPs in a-SiO$_2$ are much shorter ($\approx$ 1 nm \cite{yang2009thermal}) than both the spot radius and the nanoline widths. As such, thermal transport remains in the diffusive regime, allowing us to assess if the model can accurately capture variations in phase lag arising from changes in spot radius and nanoline width.
An analysis was performed to determine the sensitivity of each parameter used in the analytical solution. 
The sensitivity of a parameter, $S_\phi$, to the phase lag measured in FDTR is quantified by

\begin{equation} \label{eq:1}
S_{\phi} = \frac{d \ln (\phi)}{d \ln (s)} = \frac{d\phi}{ds} \times\frac{s}{\phi},
\end{equation}
which shows the normalized phase change per normalized unit change in the parameter.\cite{gundrum_thermal_2005}
A high sensitivity of a fitting parameter indicates that FDTR can be used to accurately measure that parameter. 
If the phase lag is sensitive to an input parameter, then that parameter needs to be externally quantified with low uncertainty to minimize its unintentional influence on the fitted parameters. 

The nominal parameters used in the sensitivity analyses for a-SiO$_2$ and $c$-sapphire are presented in Table~\ref{sen_MC_parameters}. The thermal conductivity of the Au-Cr nanoline was initially determined from a witness sample using the four-point probe method. The witness sample was prepared by depositing the Au-Cr layer on a separate a-SiO$_2$ substrate simultaneously with the Au nanoline deposition to ensure identical thickness and deposition conditions. The Wiedemann-Franz law was applied to convert the measured electrical conductivity to the Au-Cr thermal conductivity.
The volumetric heat capacity and the thermal conductivity of the a-SiO$_2$ substrate were obtained from the literature.~\cite{bergman_fundamentals_2017, vdi_e_v_vdi_2010} 

\renewcommand{\arraystretch}{1.3}
\setlength{\tabcolsep}{12pt} 
\begin{table}[t]
    \caption{Parameters used in the sensitivity analysis and the Monte Carlo uncertainty quantification. $k_x = k_y = k_z$ for both substrates.}
    \begin{center}
        \begin{tabular}{cccc}
        \hline
    Property & Au & a-SiO$_2$ & $c$-sapphire\\
            \hline
     $C_v$ (MJ/m$^2$-K) & 2.43 $\pm$ 0.03\cite{bergman_fundamentals_2017} & 1.58 $\pm$ 0.02\cite{vdi_e_v_vdi_2010} & 2.99 $\pm$ 0.03\cite{bergman_fundamentals_2017} \\
     {\color{black}$k_x$ (W/m-K)} & {\color{black}0} & {\color{black}1.4 $\pm$ 0.1\cite{vdi_e_v_vdi_2010}} & {\color{black}Fitted} \\
     {\color{black}$k_y$ (W/m-K)} & {\color{black}Fitted} & {\color{black}1.4 $\pm$ 0.1\cite{vdi_e_v_vdi_2010}} & {\color{black}Fitted} \\
     {\color{black}$k_z$ (W/m-K)} & {\color{black}160 $\pm$ 8$^{\mathrm{a}}$} & {\color{black}1.4 $\pm$ 0.1\cite{vdi_e_v_vdi_2010}} & {\color{black}Fitted} \\
     $d$ & 75.0 $\pm$ 1.3 nm$^{\mathrm{b}}$ & 300 $\pm$ 3 $\mu$m & 400 $\pm$ 4 $\mu$m \\
         {\color{black}$r$ ($\mu$m)}
& {\color{black}-}
& {\color{black}\makecell{4.1 $\pm$ 0.2 (10$\times$) \\ 2.3 $\pm$ 0.1 (20$\times$)}}
& {\color{black}\makecell{5.0 $\pm$ 0.2 (10$\times$) \\ 2.5 $\pm$ 0.1 (20$\times$)}}
\\
     \multicolumn{4}{l}{\color{black}$^{\mathrm{a}}$ Measured with four-point probe method on a witness sample}\\
     \multicolumn{4}{l}{$^{\mathrm{b}}$ Measured with profilometer, see Sec. S1 in Supplementray Information ($\pm$ root mean square roughness)}
        \end{tabular}
        \label{sen_MC_parameters}
    \end{center}
\end{table}

\begin{figure}[h]
 \centering
 \includegraphics[width=\linewidth]{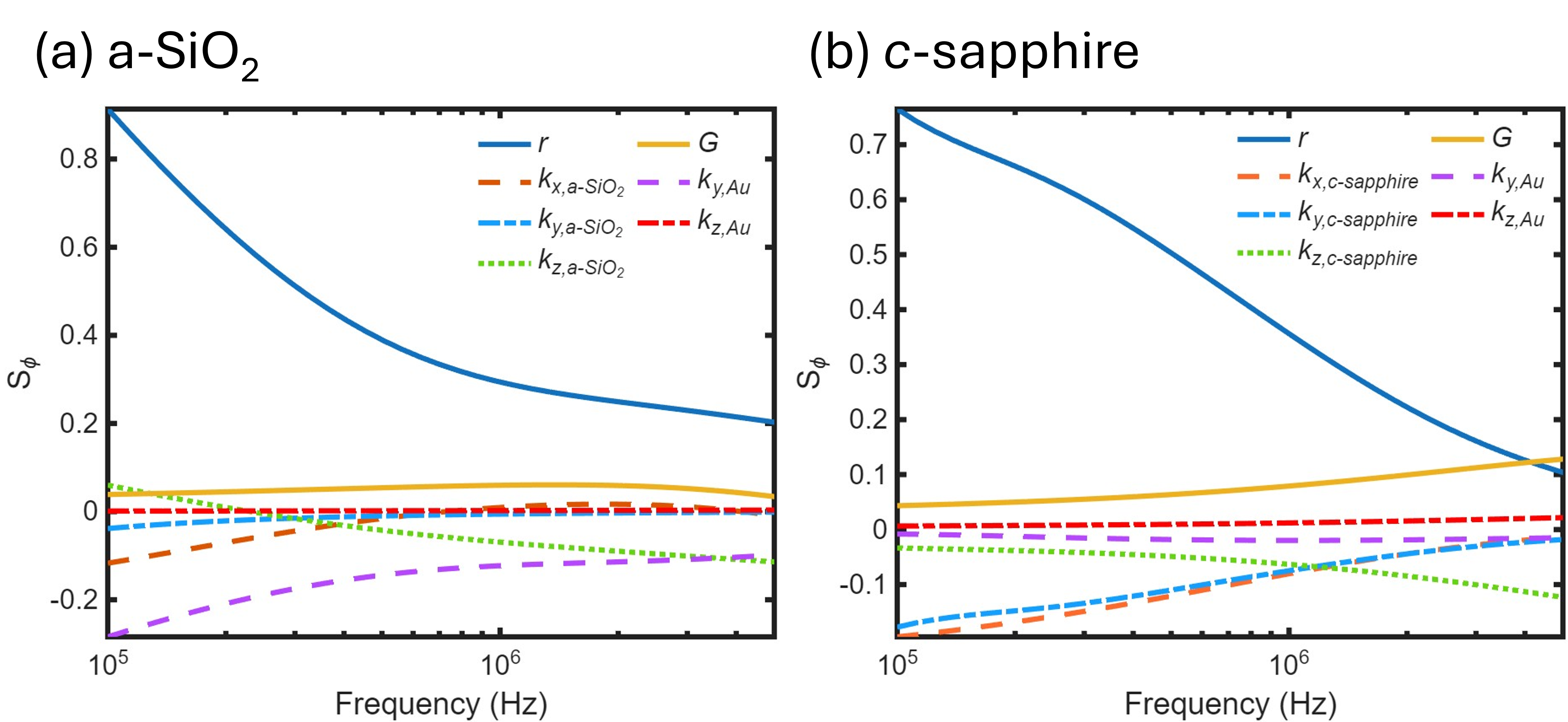}
 \caption{Sensitivity analysis for \textbf{(a)} nanolines deposited on the a-SiO$_2$ substrate and \textbf{(b)} nanolines deposited on the $c$-sapphire substrate.}
 \label{sensitivity_plot}
\end{figure}

\begin{figure}[h]
 \centering
 \includegraphics[width=\linewidth]{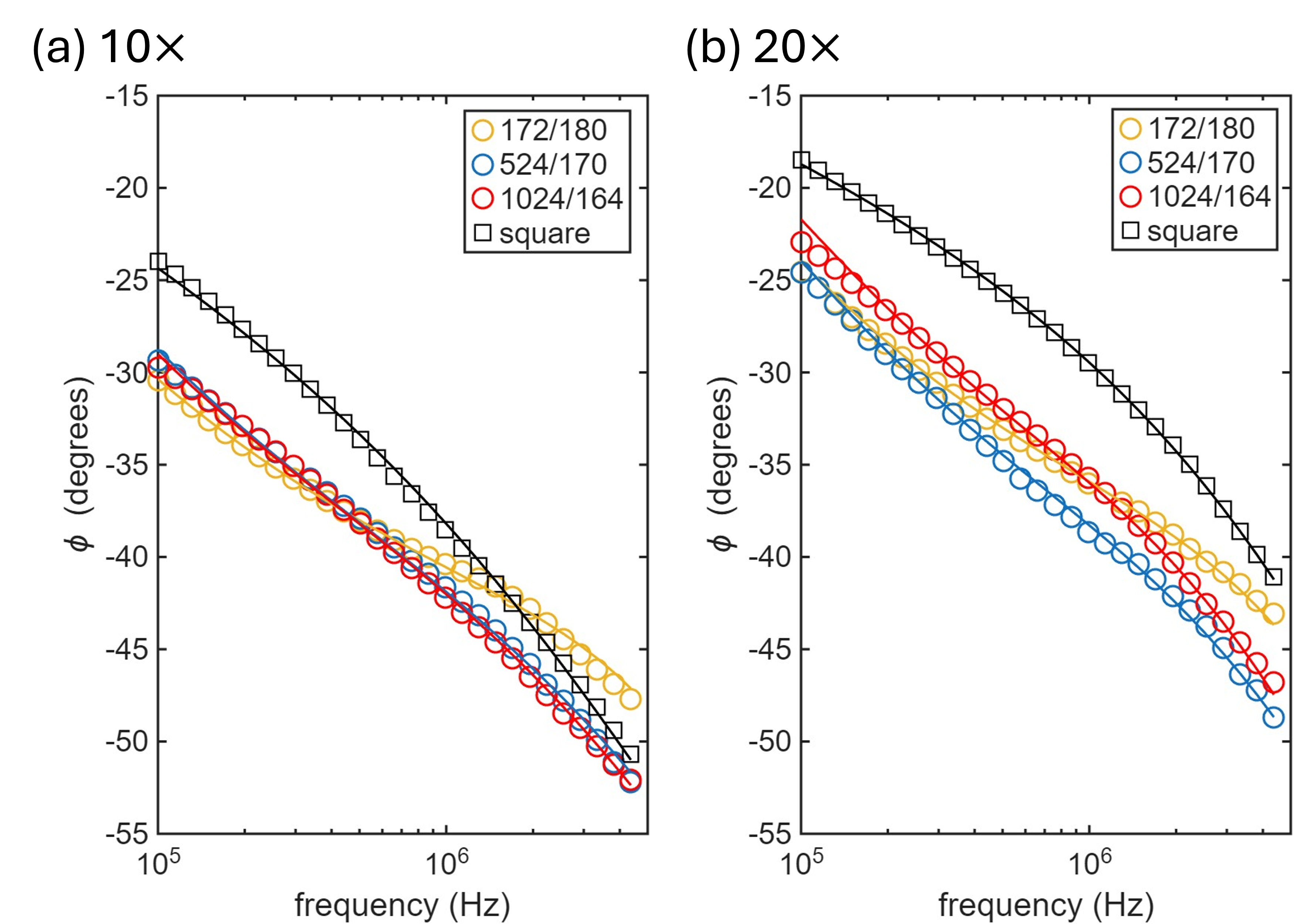}
 \caption{Phase lag and analytical solution fits for nanolines deposited on a-SiO$_2$ with \textbf{(a)} 10$\times$ objective ($r = 4.1 \pm 0.2 $ $\mu$m) and \textbf{(b)} 20$\times$ objective ($r = 2.3 \pm 0.1 $ $\mu$m). The legend uses the notation $w$/gap (nm); e.g., 172/180 corresponds to a width $w$ of 172 nm and a gap ($L-w)$ of 180 nm.}
 \label{fitting_fused_silica}
\end{figure}

\begin{figure}[h]
 \centering
 \includegraphics[width =\linewidth]{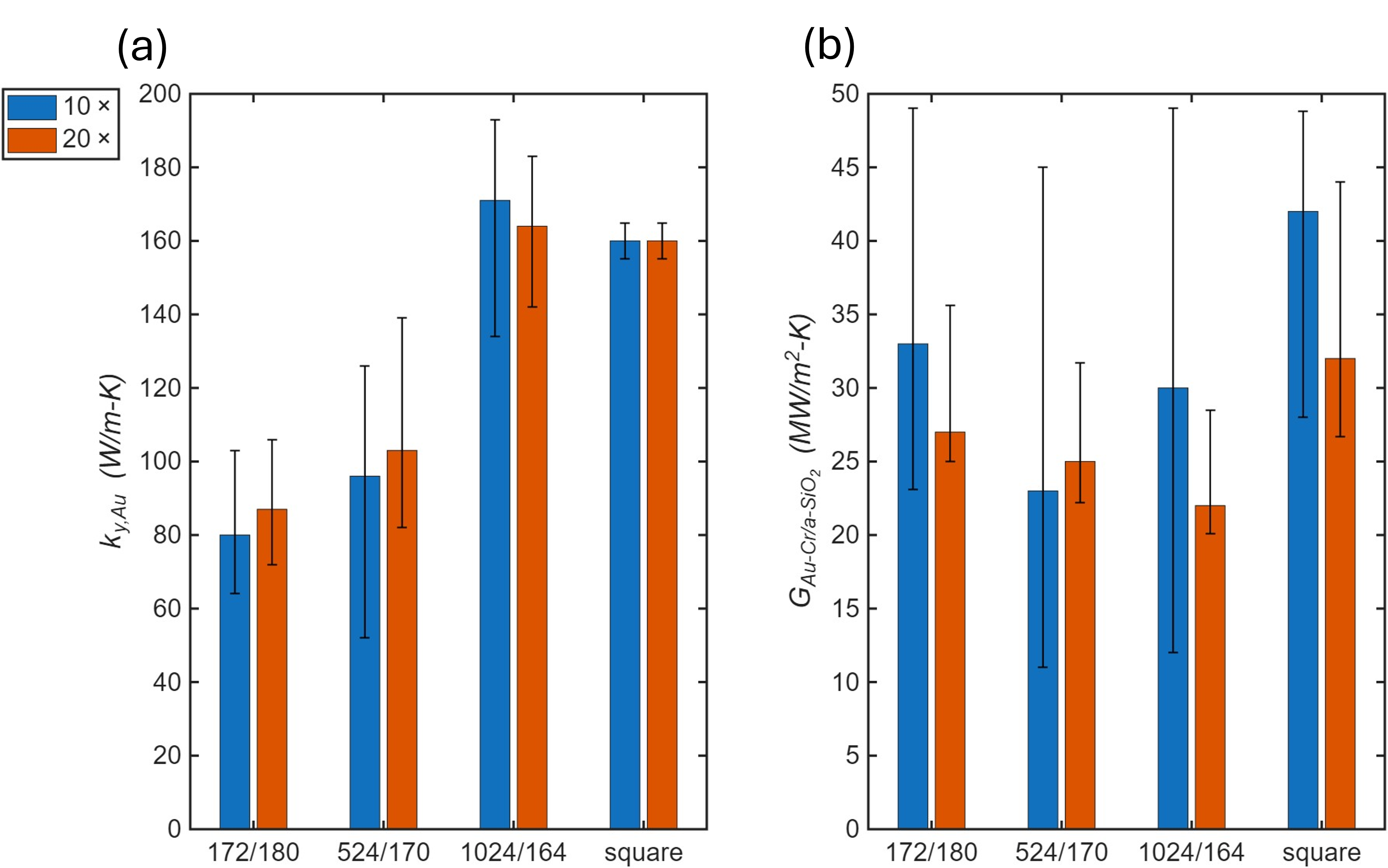}
 \caption{Extracted \textbf{(a)} $k_{y,Au}$ and \textbf{(b)} $G_{Au-Cr/a-SiO_2}$ for nanolines deposited on a-SiO$_2$ measured with 10$\times$ objective ($r = 4.1 \pm 0.2 $ $\mu$m) and 20$\times$ objective ($r = 2.3 \pm 0.1 $ $\mu$m). The legend uses the notation $w$/gap (nm); e.g., 172/180 corresponds to a width $w$ of 172 nm and a  gap ($L-w)$ of 180 nm.}
 \label{extracted_fused_silica}
\end{figure}

The a-SiO$_2$ sensitivity analysis is shown in Fig.~\ref{sensitivity_plot}(a) and indicates sensitivity to $k_{y,Au}$ and $G_{Au-Cr/a-SiO_2}$. The initial fit using $k_{y,Au}$ from the witness sample was unsuccessful, suggesting a possible variation in $k_{y,Au}$ with different nanoline widths. We therefore simultaneously extracted $k_{y,Au}$ and $G_{Au-Cr/a-SiO_2}$ for three nanoline patterns and the square patches for each laser spot radius. 
The analysis is insensitive to the cross-plane thermal conductivity of the Au nanolines ($k_{z,Au}$), and therefore the value measured from the witness sample was used and treated as a fixed parameter.
The analysis is also highly sensitive to the laser spot radius [Fig. \ref{sensitivity_plot}(a)]. {\color{black}The uncertainty of the laser spot size was characterized with previous knife-edge measurements and the reference sample (Au on a-SiO$_2$) measurement.} {\color{black}As verified by 3D Finite Element Method (FEM) simulations (Fig. S7), the laser spot position relative to the nanolines has a negligible effect on the measured phase. The simulations considered a lateral offset of 100 nm from the nanoline center, corresponding to twice the resolution of the micro-manipulator.}
The analytical fits to the measured phase lag are plotted for laser spot radii of 4.1 {\color{black}$\pm$ 0.2 $\mu$m} [Fig.~\ref{fitting_fused_silica}(a)] and 2.3 {\color{black}$\pm$ 0.1 $\mu$m} [Fig.~\ref{fitting_fused_silica}(b)]. 
The values of $k_{y,Au}$ and $G_{Au-Cr/a-SiO_2}$ extracted from these fits are plotted in Figs.~\ref{extracted_fused_silica}(a) and \ref{extracted_fused_silica}(b). 
The uncertainty of the fitting parameters was quantified using a Monte Carlo method. \cite{yang_uncertainty_2016}
For the input parameters listed in Table~\ref{sen_MC_parameters}, we assumed Gaussian distributions using their estimated uncertainties as standard deviations. 
In each Monte Carlo iteration, the input parameters were randomly sampled from each distribution, and the unknown parameters were extracted by fitting the newly constructed analytical solution to the experimental data. 
This procedure was repeated for 1000 iterations to generate distributions of the fitted parameters. Uncertainties are reported as the central 95$\%$ confidence interval to account for possible skewness of the distribution.

The results in Figs.~\ref{fitting_fused_silica}(a) and \ref{fitting_fused_silica}(b) show that the measured phase lag in FDTR changes upon changes in the nanoline pattern geometry [variations within (a) or (b)], as well as upon changes in the laser spot radius [variations across (a) and (b) for the same nanoline pattern]. 
The change in phase lag due to a change in nanoline dimensions was previously measured using TDTR and modeled using an infinite spot radius approximation.\cite{Minnich_2011,Minnich_2013,hu2015spectral,chen2018quasiballistic}
Comparing across Figs.~\ref{fitting_fused_silica}(a) and \ref{fitting_fused_silica}(b), the need to incorporate the effect of the finite laser spot radius is evident.
There is an increase in phase lag (i.e., a larger negative value) recorded for the larger spot radius for the same nanoline pattern, which is consistent with the effect of using a larger spot radius in the conventional FDTR heat diffusion model.~\cite{wang_reducing_2022,malen_optical_2011} The laser spot radius dependence is further explored in Appendix B, where we emphasize differences relative to the infinite spot radius assumption.

As shown in Fig.~\ref{extracted_fused_silica}, we find that $k_{y,Au}$ decreases by roughly half as the nanoline width is reduced from 1024 nm to 172 nm for both spot radii, based on our set of three measured linewidths. We hypothesize that this reduction is due to a modified grain structure in the Au when it is deposited between the narrow {\color{black}e-beam resist (AR-P 6200.11, CSAR) walls, as noted in Sec. S1.} In the most extreme case, the Au is deposited in a 172 nm wide trench between 300 nm tall CSAR walls. $k_{y,Au}$ for the nanoline width of 1024 nm is comparable to that for the square patch, suggesting that the grain structure is less affected at this larger width.
$G_{Au-Cr/a-SiO_2}$ for all patterns are comparable to the $G_{Au-Cr/a-SiO_2}$ extracted from the square patch within the experimental uncertainty. These values also fall within the range we routinely obtain (20-60 MW/m$^2$-K) for this interface from reference measurements used to calibrate the laser spot radius prior to experiments.

\subsection{Sapphire}\label{subsec:sapphireresults}
\begin{figure}[h]
 \centering
 \includegraphics[width = \linewidth]{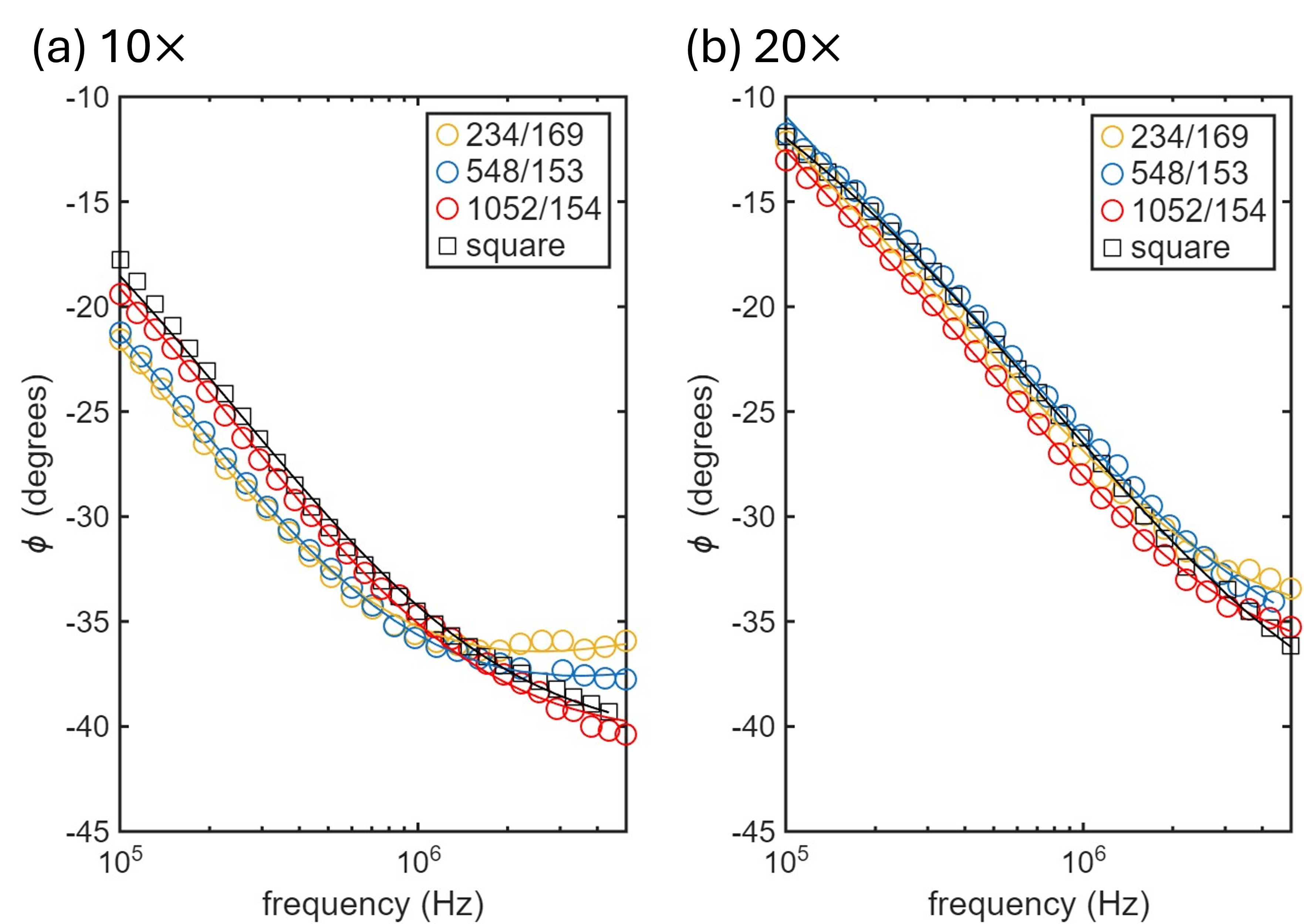}
 \caption{Phase lag and analytical solution fits for nanolines deposited on the $c$-sapphire measured with \textbf{(a)} 10$\times$ objective ($r = 5.0 \pm 0.2 $ $\mu$m) and \textbf{(b)} 20$\times$ objective ($r = 2.5 \pm 0.1 $ $\mu$m). The legend uses the notation $w$/gap (nm); e.g., 234/169 corresponds to a width $w$ of 234 nm and a gap ($L-w$) of 169 nm.}
 \label{fitting_sapphire}
\end{figure}

We next studied $c$-sapphire to further experimentally validate the analytical solution within a diffusive thermal transport regime. FDTR phase lag data and fits for the Au nanolines on $c$-sapphire are shown in Figs.~\ref{fitting_sapphire}(a) (10$\times$ objective) and \ref{fitting_sapphire}(b) (20$\times$ objective). We co-fit the thermal boundary conductance at the Au-Cr/$c$-sapphire interface ($G_{Au-Cr/c-sapphire}$) and the thermal conductivity of $c$-sapphire ($k_{c-sapphire}$) for the different nanoline patterns. The extracted $k_{y,Au}$ from the Au nanolines on a-SiO$_2$ was used as a fixed parameter. Because $k_{y,Au}$ varies with the nanoline width on a-SiO$_2$, we selected the $k_{y,Au}$ value obtained from the nanolines deposited on a-SiO$_2$ using the 20$\times$ objective with the closest linewidth. We note that FDTR has a much lower sensitivity to $k_{y,Au}$ for the $c$-sapphire substrate compared to the a-SiO$_2$ substrate, as shown in Fig.~\ref{sensitivity_plot}(b). 

\begin{figure}[h]
 \centering
 \includegraphics[width=\linewidth]{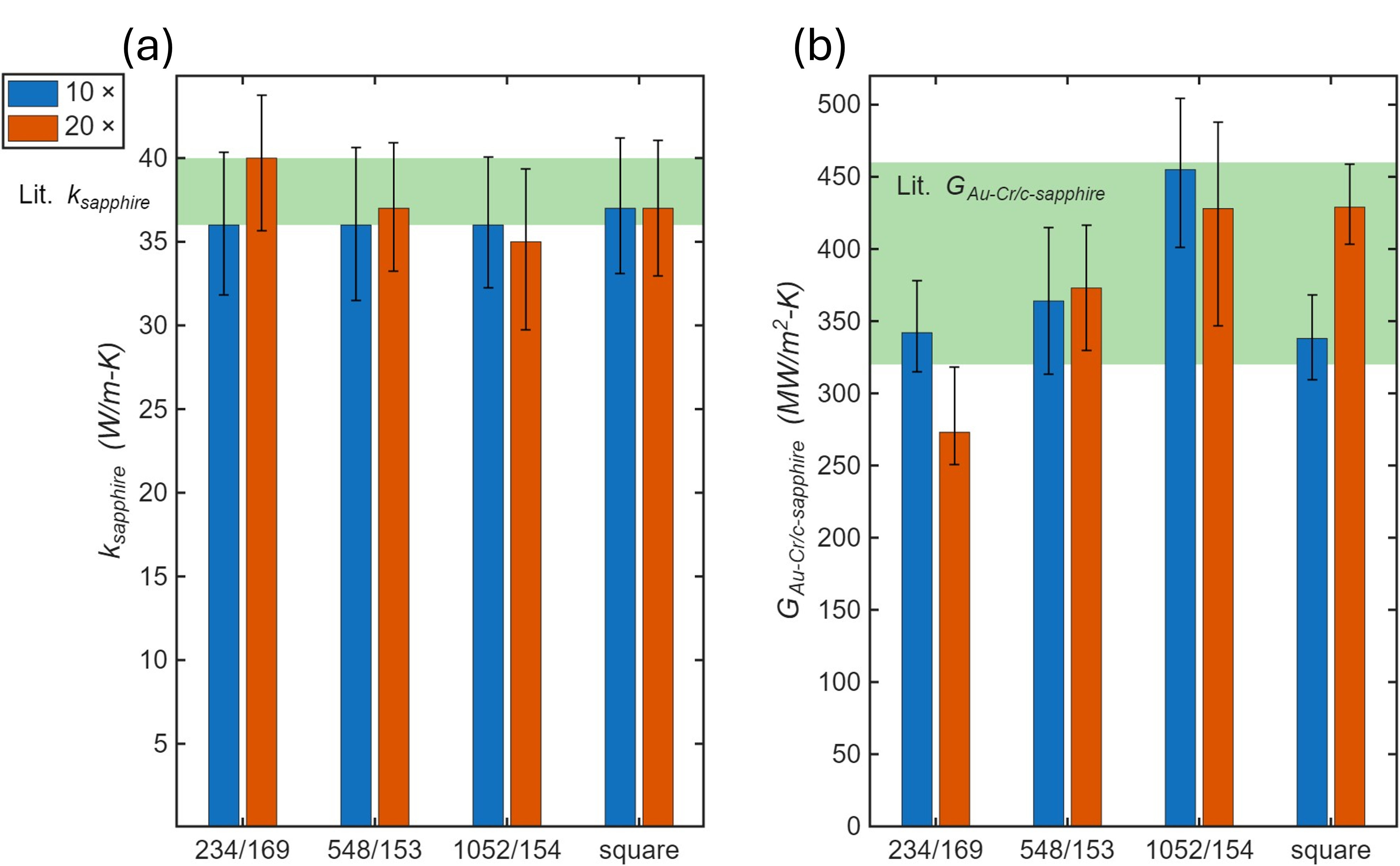}
 \caption{Extracted \textbf{(a)} $k_{c-sapphire}$ and \textbf{(b)} $G_{Au-Cr/c-sapphire}$ for nanolines deposited on the $c$-sapphire measured with 10$\times$ objective ($r = 5.0 \pm 0.2 $ $\mu$m) and 20$\times$ objective ($r = 2.5 \pm 0.1 $ $\mu$m). The legend uses the notation $w$/gap (nm); e.g., 234/169 corresponds to a width $w$ of 234 nm and a gap ($L-w)$ of 169 nm. The shaded lines represent the literature-reported $k_{c-sapphire}$ and $G_{Au-Cr/c-sapphire}$, which are $38 \pm 2$ W/m-K~\cite{collins_examining_2014} and $390 \pm 70$ MW/m$^2$-K.~\cite{jeong_enhancement_2016}}
 \label{extracted_sapphire}
\end{figure}

The fitted analytical solution plotted in Figs.~\ref{fitting_sapphire}(a) and \ref{fitting_sapphire}(b) shows excellent agreement with the measured phase lag data for both laser spot radii.
The extracted $k_{c-sapphire}$ for all patterns plotted in Fig.~\ref{extracted_sapphire}(a) are within the error margins of $k_{c-sapphire}$ from the square patch as well as the reported bulk value of 38 $\pm$ 2 W/m-K.~\cite{collins_examining_2014} These results indicate that the thermal conductivity suppression due to ballistic phonon transport is insignificant within the range of MFPs we explored with nanolines, as approximately 90 percent of the bulk $k_{c-sapphire}$ is accumulated from MFPs less than 100 nm.~\cite{zhou_effective_2026} As shown in Fig.~\ref{extracted_sapphire}(b), $G_{Au-Cr/c-sapphire}$ from all patterns is comparable to the values extracted from the square pattern as well as to a literature-reported value (390 $\pm$ 70 MW/m$^2$-K).\cite{jeong_enhancement_2016} Variation in the extracted $G_{Au-Cr/c-sapphire}$ is likely due to a relatively low sensitivity to $G_{Au-Cr/c-sapphire}$, as shown in Fig.~\ref{sensitivity_plot}(b).

\section{Conclusion}
We developed and validated a nanoline-based FDTR technique with a finite spot radius that targets studies of $k_{accum}$ without the need for samples of varied thickness. A new analytical solution to the heat diffusion equation was developed to account for finite Gaussian laser spots and nanoline periodicity. To validate the solution, we presented FDTR measurements of Au nanolines  with widths ranging from 100 nm to 1000 nm that were deposited on a-SiO$_2$ and $c$-sapphire substrates, which exhibit diffusive heat transport due to their short energy carrier MFPs. Measurements on both a-SiO$_2$ and sapphire substrates showed good agreement between the experimental phase lag data and the analytical model, validating the robustness of the approach. 
The high sensitivity of $k_{y,Au}$ on a-SiO$_2$ highlights the importance of performing sensitivity analysis for each new set of experimental conditions. {\color{black}The present framework can be extended beyond one-dimensional nanoline gratings to two-dimensional periodic arrays, such as square or rectangular metallic features, by incorporating an additional Fourier series expansion in the transverse direction.} Moving forward, the established FDTR framework with its commonly used range of spot radii can now be extended to measure phonon MFP contributions to $k_{accum}$ across a wide range spanning sub-micron to micrometer length scales.

\section*{Author's Contributions}
\noindent\textbf{Abhishek Pathak}: Conceptualization (lead); Data curation (equal); Formal analysis (equal); Methodology (lead); Software (lead); Visualization (equal); Validation (equal); Writing - original draft (equal); Writing - review $\&$ editing (equal). \textbf{Seungwon Park}: Conceptualization (equal); Data curation (equal); Formal analysis (equal); Methodology (equal); Software (equal); Visualization (equal); Validation (equal); Writing - original draft (equal); Writing - review $\&$ editing (equal). \textbf{Shubham Deshmukh}: Formal analysis (support); Methodology (support); Software (support);  Validation (support); Writing - original draft (support); Writing - review $\&$ editing (support). \textbf{Alan J. H. McGaughey}: Funding acquisition (lead); Project administration (lead); Supervision (lead); Writing - review $\&$ editing (lead). \textbf{Jonathan A. Malen}: Funding acquisition (lead); Project administration (lead); Supervision (lead); Writing - review $\&$ editing (lead).

\section*{Acknowledgments}
This work was supported by the Army Research Office Award No. W911NF2220191. The authors acknowledge the use of the Bertucci Nanotechnology Laboratory at Carnegie Mellon University supported by grant BNL-78657879 and use of the Materials Characterization Facility at Carnegie Mellon University supported by grant MCF-677785.
\section*{Conflict of Interest}
The authors have no conflicts to disclose.

\section*{Data Availability}
The data that support the findings of this study are available from the corresponding author upon reasonable request.

\bibliography{ref}

@article{frazer2019engineering,
  title={Engineering nanoscale thermal transport: Size-and spacing-dependent cooling of nanostructures},
  author={Frazer, Travis D and Knobloch, Joshua L and Hoogeboom-Pot, Kathleen M and Nardi, Damiano and Chao, Weilun and Falcone, Roger W and Murnane, Margaret M and Kapteyn, Henry C and Hernandez-Charpak, Jorge N},
  journal={Physical Review Applied},
  volume={11},
  number={2},
  pages={024042},
  year={2019},
  publisher={APS}
}

@article{hoogeboom2015new,
  title={A new regime of nanoscale thermal transport: Collective diffusion increases dissipation efficiency},
  author={Hoogeboom-Pot, Kathleen M and Hernandez-Charpak, Jorge N and Gu, Xiaokun and Frazer, Travis D and Anderson, Erik H and Chao, Weilun and Falcone, Roger W and Yang, Ronggui and Murnane, Margaret M and Kapteyn, Henry C and others},
  journal={Proceedings of the National Academy of Sciences},
  volume={112},
  number={16},
  pages={4846--4851},
  year={2015},
  publisher={National Academy of Sciences}
}

@article{siemens2010quasi,
  title={Quasi-ballistic thermal transport from nanoscale interfaces observed using ultrafast coherent soft X-ray beams},
  author={Siemens, Mark E and Li, Qing and Yang, Ronggui and Nelson, Keith A and Anderson, Erik H and Murnane, Margaret M and Kapteyn, Henry C},
  journal={Nature materials},
  volume={9},
  number={1},
  pages={26--30},
  year={2010},
  publisher={Nature Publishing Group UK London}
}

@article{cahill2004analysis,
  title={Analysis of heat flow in layered structures for time-domain thermoreflectance},
  author={Cahill, David G and others},
  journal={Review of scientific instruments},
  volume={75},
  number={12},
  pages={5119},
  year={2004}
}

@article{schmidt2009frequency,
  title={A frequency-domain thermoreflectance method for the characterization of thermal properties},
  author={Schmidt, Aaron J and Cheaito, Ramez and Chiesa, Matteo},
  journal={Review of scientific instruments},
  volume={80},
  number={9},
  year={2009},
  publisher={AIP Publishing}
}

@article{kim2021origin,
  title={Origin of micrometer-scale propagation lengths of heat-carrying acoustic excitations in amorphous silicon},
  author={Kim, Taeyong and Moon, Jaeyun and Minnich, Austin J},
  journal={Physical Review Materials},
  volume={5},
  number={6},
  pages={065602},
  year={2021},
  publisher={APS}
}

@article{cuffe2015reconstructing,
  title={Reconstructing phonon mean-free-path contributions to thermal conductivity using nanoscale membranes},
  author={Cuffe, John and Eliason, Jeffrey K and Maznev, Alexei A and Collins, Kimberlee C and Johnson, Jeremy A and Shchepetov, Andrey and Prunnila, Mika and Ahopelto, Jouni and Sotomayor Torres, Clivia M and Chen, Gang and others},
  journal={Physical Review B},
  volume={91},
  number={24},
  pages={245423},
  year={2015},
  publisher={APS}
}

@article{johnson2013direct,
  title={Direct Measurement of Room-Temperature Nondiffusive Thermal Transport Over Micron Distances in a Silicon Membrane},
  author={Johnson, Jeremy A and Maznev, AA and Cuffe, John and Eliason, Jeffrey K and Minnich, Austin J and Kehoe, Timothy and Torres, Clivia M Sotomayor and Chen, Gang and Nelson, Keith A},
  journal={Physical review letters},
  volume={110},
  number={2},
  pages={025901},
  year={2013},
  publisher={APS}
}

@article{regner2015interpretation,
  title={Interpretation of thermoreflectance measurements with a two-temperature model including non-surface heat deposition},
  author={Regner, Keith T and Wei, Lien Chin and Malen, Jonathan A},
  journal={Journal of Applied Physics},
  volume={118},
  number={23},
  year={2015},
  publisher={AIP Publishing}
}

@article{regner2014analytical,
  title={Analytical interpretation of nondiffusive phonon transport in thermoreflectance thermal conductivity measurements},
  author={Regner, Keith T and McGaughey, Alan JH and Malen, Jonathan A},
  journal={Phys. Rev. B},
  volume={90},
  number={6},
  pages={064302},
  year={2014}
}

@manual{COMSOL,
  title={{Wave Optics Module}},
  author={{COMSOL AB}},
  organization={{COMSOL AB}},
  year={{2025}},
  edition={{COMSOL Multiphysics\textregistered{} v. 6.4}}, 
  address={{Stockholm, Sweden}},
  url={{https://www.comsol.com}} 
}

@article{polyanskiy2024refractiveindex,
  title={Refractiveindex. info database of optical constants},
  author={Polyanskiy, Mikhail N},
  journal={Scientific Data},
  volume={11},
  number={1},
  pages={94},
  year={2024},
  publisher={Nature Publishing Group UK London}
}

@article{maznev2011onset,
  title={Onset of nondiffusive phonon transport in transient thermal grating decay},
  author={Maznev, Alexei A and Johnson, Jeremy A and Nelson, Keith A},
  journal={Physical Review B—Condensed Matter and Materials Physics},
  volume={84},
  number={19},
  pages={195206},
  year={2011},
  publisher={APS}
}

@article{yang2009thermal,
  title={Thermal transport in CO2 laser irradiated fused silica: In situ measurements and analysis},
  author={Yang, Steven T and Matthews, Manyalibo J and Elhadj, Selim and Draggoo, Vaughn G and Bisson, Scott E},
  journal={Journal of Applied Physics},
  volume={106},
  number={10},
  year={2009},
  publisher={AIP Publishing}
}

@article{zhang2025probing,
  title={Probing Phonon Mean-Free-Path Distribution via Thickness-Dependent Thermal Conductivity in Epitaxial SrSnO3},
  author={Zhang, Chi and Liu, Fengdeng and Kim, Donghwan and Xu, Xiaotian and Guo, Silu and Pei, Yankai and Mkhoyan, K Andre and Feng, Tianli and Jalan, Bharat and Wang, Xiaojia},
  journal={Materials Today Physics},
  pages={101991},
  year={2025},
  publisher={Elsevier}
}

@article{schafer1980disordered,
  title={Disordered system with n orbitals per site: Lagrange formulation, hyperbolic symmetry, and Goldstone modes},
  author={Sch{\"a}fer, Lothar and Wegner, Franz},
  journal={Zeitschrift f{\"u}r Physik B Condensed Matter},
  volume={38},
  number={2},
  pages={113--126},
  year={1980},
  publisher={Springer}
}

@article{zeng2015measuring,
  title={Measuring phonon mean free path distributions by probing quasiballistic phonon transport in grating nanostructures},
  author={Zeng, Lingping and Collins, Kimberlee C and Hu, Yongjie and Luckyanova, Maria N and Maznev, Alexei A and Huberman, Samuel and Chiloyan, Vazrik and Zhou, Jiawei and Huang, Xiaopeng and Nelson, Keith A and others},
  journal={Scientific reports},
  volume={5},
  number={1},
  pages={17131},
  year={2015},
  publisher={Nature Publishing Group UK London}
}

@article{minnich2011thermal,
  title={Thermal conductivity spectroscopy technique to measure phonon mean free paths},
  author={Minnich, Austin J and Johnson, Jeremiah A and Schmidt, Aaron Jerome and Esfarjani, Keivan and Dresselhaus, Mildred Spiewak and Nelson, Keith A and Chen, Gang},
  journal={Physical review letters},
  volume={107},
  number={9},
  pages={095901},
  year={2011},
  publisher={APS}
}

@article{braun2017upper,
  title={Upper limit to the thermal penetration depth during modulated heating of multilayer thin films with pulsed and continuous wave lasers: A numerical study},
  author={Braun, Jeffrey L and Hopkins, Patrick E},
  journal={Journal of Applied Physics},
  volume={121},
  number={17},
  year={2017},
  publisher={AIP Publishing}
}

@article{regner2013broadband,
  title={Broadband phonon mean free path contributions to thermal conductivity measured using frequency domain thermoreflectance},
  author={Regner, Keith T and Sellan, Daniel P and Su, Zonghui and Amon, Cristina H and McGaughey, Alan JH and Malen, Jonathan A},
  journal={Nature communications},
  volume={4},
  number={1},
  pages={1640},
  year={2013},
  publisher={Nature Publishing Group UK London}
}

@article{koh2007frequency,
  title={Frequency dependence of the thermal conductivity of semiconductor alloys},
  author={Koh, Yee Kan and Cahill, David G},
  journal={Physical Review B—Condensed Matter and Materials Physics},
  volume={76},
  number={7},
  pages={075207},
  year={2007},
  publisher={APS}
}

@article{negi2023thickness,
  title={Thickness-Dependent Thermal Conductivity and Phonon Mean Free Path Distribution in Single-Crystalline Barium Titanate},
  author={Negi, Ankit and Rodriguez, Alejandro and Zhang, Xuanyi and Comstock, Andrew H and Yang, Cong and Sun, Dali and Jiang, Xiaoning and Kumah, Divine and Hu, Ming and Liu, Jun},
  journal={Advanced Science},
  volume={10},
  number={19},
  pages={2301273},
  year={2023},
  publisher={Wiley Online Library}
}

@misc{carslaw1959conduction,
  title={Conduction of heat in solids, Clarendon},
  author={Carslaw, HS and Jaeger, JC},
  year={1959},
  publisher={Oxford}
}

@article{chen2018quasiballistic,
  title={Quasiballistic thermal transport from nanoscale heaters and the role of the spatial frequency},
  author={Chen, Xiangwen and Hua, Chengyun and Zhang, Hang and Ravichandran, Navaneetha K and Minnich, Austin J},
  journal={Physical Review Applied},
  volume={10},
  number={5},
  pages={054068},
  year={2018},
  publisher={APS}
}

@article{hu2015spectral,
  title={Spectral mapping of thermal conductivity through nanoscale ballistic transport},
  author={Hu, Yongjie and Zeng, Lingping and Minnich, Austin J and Dresselhaus, Mildred S and Chen, Gang},
  journal={Nature nanotechnology},
  volume={10},
  number={8},
  pages={701--706},
  year={2015},
  publisher={Nature Publishing Group UK London}
}

@article{Minnich_2013, title={MEASURING PHONON MEAN FREE PATHS USING THERMAL CONDUCTIVITY SPECTROSCOPY}, volume={16}, ISSN={1049-0787, 2375-0294}, url={https://www.dl.begellhouse.com/references/5756967540dd1b03,6950786d25f970a1,62afdbe346b48f01.html}, DOI={10.1615/AnnualRevHeatTransfer.v16.70}, abstractNote={Size effects in heat conduction, which occur when phonon mean free paths (MFPs) are comparable to characteristic lengths, are being extensively explored in many...}, journal={Annual Review of Heat Transfer}, publisher={Begel House Inc.}, author={Minnich, Austin J.}, year={2013}}

@phdthesis{Minnich_2011, type={Thesis}, title={Exploring electron and phonon transport at the nanoscale for thermoelectric energy conversion}, rights={M.I.T. theses are protected by  copyright. They may be viewed from this source for any purpose, but  reproduction or distribution in any format is prohibited without written  permission. See provided URL for inquiries about permission.}, url={https://dspace.mit.edu/handle/1721.1/67593}, abstractNote={Thermoelectric materials are capable of solid-state direct heat to electricity energy conversion and are ideal for waste heat recovery applications due to their simplicity, reliability, and lack of environmentally harmful working fluids. Recently, nanostructured thermoelectrics have demonstrated remarkably enhanced energy conversion efficiencies, primarily due to a reduction in lattice thermal conductivity. Despite these advances, much remains unknown about heat transport in these materials, and further efficiency improvements will require a detailed understanding of how the heat carriers, electrons and phonons, are affected by nanostructures. To elucidate these processes, in this thesis we investigate nanoscale transport using both modeling and experiment. The first portion of the thesis studies how electrons and phonons are affected by grain boundaries in nanocomposite thermoelectric materials, where the grain sizes are smaller than mean free paths (MFPs). We use the Boltzmann transport equation (BTE) and a new grain boundary scattering model to understand how thermoelectric properties are affected in nanocomposites, as well as to identify strategies which could lead to more efficient materials. The second portion of the thesis focuses on determining how to more directly measure heat carrier properties like frequency-dependent MFPs. Knowledge of phonon MFPs is crucial to understanding and engineering nanoscale transport, yet MFPs are largely unknown even for bulk materials and few experimental techniques exist to measure them. We show that performing macroscopic measurements cannot reveal the MFPs; instead, we must study transport at the scales of the MFPs, in the quasi- ballistic transport regime. To investigate transport at these small length scales, we first numerically solve the frequency-dependent phonon BTE, which is valid even in the absence of local thermal equilibrium, unlike heat diffusion theory. Next, we introduce a novel thermal conductivity spectroscopy technique which can measure MFP distributions over a wide range of length scales and materials using observations of quasi-ballistic heat transfer in a pump-probe experiment. By observing the changes in thermal resistance as a heated area size is systematically varied, the thermal conductivity contributions from different MFP phonons can be determined. We present the first experimental measurements of the MFP distribution in silicon at cryogenic temperatures. Finally, we develop a modification of this technique which permits us to study transport at scales much smaller than the diffraction limit of approximately one micron. It is important to access these length scales as many technologically relevant materials like thermoelectrics have MFPs in the deep submicron regime. To beat the diffraction limit, we use electron-beam lithography to pattern metallic nano dot arrays with diameters in the hundreds of nanometers range. Because the effective length scale for heat transfer is the dot diameter rather than the optical beam diameter, we are able to study nanoscale heat transfer while still achieving ultrafast time resolution. We demonstrate the modified technique by measuring the MFP distribution in sapphire. Considering the crucial importance of the knowledge of MFPs to understanding and engineering nanoscale transport, we expect these newly developed techniques to be useful for a variety of energy applications, particularly for thermoelectrics, as well as for gaining a fundamental understanding of nanoscale heat transport.}, note={Accepted: 2011-12-09T21:29:23Z}, school={Massachusetts Institute of Technology}, author={Minnich, Austin Jerome}, year={2011}}

@article{zeng_measuring_2015,
	title = {Measuring {Phonon} {Mean} {Free} {Path} {Distributions} by {Probing} {Quasiballistic} {Phonon} {Transport} in {Grating} {Nanostructures}},
	volume = {5},
	issn = {2045-2322},
	url = {https://www.nature.com/articles/srep17131},
	doi = {10.1038/srep17131},
	number = {1},
	urldate = {2025-10-03},
	journal = {Scientific Reports},
	author = {Zeng, Lingping and Collins, Kimberlee C. and Hu, Yongjie and Luckyanova, Maria N. and Maznev, Alexei A. and Huberman, Samuel and Chiloyan, Vazrik and Zhou, Jiawei and Huang, Xiaopeng and Nelson, Keith A. and Chen, Gang},
	month = nov,
	year = {2015},
	pages = {17131},
}

@article{regner_advances_2015,
	title = {Advances in {Studying} {Phonon} {Mean} {Free} {Path} {Dependent} {Contributions} to {Thermal} {Conductivity}},
	volume = {19},
	issn = {1556-7265, 1556-7273},
	url = {http://www.tandfonline.com/doi/full/10.1080/15567265.2015.1045640},
	doi = {10.1080/15567265.2015.1045640},
	number = {3},
	urldate = {2025-10-03},
	journal = {Nanoscale and Microscale Thermophysical Engineering},
	author = {Regner, Keith T. and Freedman, Justin P. and Malen, Jonathan A.},
	month = jul,
	year = {2015},
	pages = {183--205},
}

@article{gundrum_thermal_2005,
	title = {Thermal conductance of metal-metal interfaces},
	volume = {72},
	copyright = {http://link.aps.org/licenses/aps-default-license},
	issn = {1098-0121, 1550-235X},
	url = {https://link.aps.org/doi/10.1103/PhysRevB.72.245426},
	doi = {10.1103/PhysRevB.72.245426},
	number = {24},
	urldate = {2024-08-24},
	journal = {Physical Review B},
	author = {Gundrum, Bryan C. and Cahill, David G. and Averback, Robert S.},
	month = dec,
	year = {2005},
	pages = {245426},
}

@article{yang_uncertainty_2016,
	title = {Uncertainty analysis of thermoreflectance measurements},
	volume = {87},
	issn = {0034-6748, 1089-7623},
	url = {https://pubs.aip.org/rsi/article/87/1/014901/367530/Uncertainty-analysis-of-thermoreflectance},
	doi = {10.1063/1.4939671},
	number = {1},
	urldate = {2024-04-23},
	journal = {Review of Scientific Instruments},
	author = {Yang, Jia and Ziade, Elbara and Schmidt, Aaron J.},
	month = jan,
	year = {2016},
	pages = {014901},
}

@article{jeong_enhancement_2016,
	title = {Enhancement of {Thermal} {Conductance} at {Metal}-{Dielectric} {Interfaces} {Using} {Subnanometer} {Metal} {Adhesion} {Layers}},
	volume = {5},
	copyright = {http://link.aps.org/licenses/aps-default-license},
	issn = {2331-7019},
	url = {https://link.aps.org/doi/10.1103/PhysRevApplied.5.014009},
	doi = {10.1103/PhysRevApplied.5.014009},
	number = {1},
	urldate = {2026-01-14},
	journal = {Physical Review Applied},
	author = {Jeong, Minyoung and Freedman, Justin P. and Liang, Hongliang Joe and Chow, Cheng-Ming and Sokalski, Vincent M. and Bain, James A. and Malen, Jonathan A.},
	month = jan,
	year = {2016},
	pages = {014009},
}

@book{vdi_e_v_vdi_2010,
	address = {Berlin, Heidelberg},
	title = {{VDI} {Heat} {Atlas}},
	copyright = {http://www.springer.com/tdm},
	isbn = {978-3-540-77876-9 978-3-540-77877-6},
	url = {https://link.springer.com/10.1007/978-3-540-77877-6},
	doi = {10.1007/978-3-540-77877-6},
	urldate = {2026-02-12},
	publisher = {Springer Berlin Heidelberg},
	editor = {{VDI e. V.}},
	year = {2010},
}

@book{bergman_fundamentals_2017,
	address = {Hoboken, NJ},
	edition = {Eighth edition},
	title = {Fundamentals of heat and mass transfer},
	isbn = {978-1-119-32042-5 978-1-119-33767-6},
	publisher = {John Wiley \& Sons},
	author = {Bergman, Theodore L. and Lavine, Adrienne S.},
	year = {2017},
}

@article{malen_optical_2011,
	title = {Optical {Measurement} of {Thermal} {Conductivity} {Using} {Fiber} {Aligned} {Frequency} {Domain} {Thermoreflectance}},
	volume = {133},
	issn = {0022-1481, 1528-8943},
	url = {https://asmedigitalcollection.asme.org/heattransfer/article/doi/10.1115/1.4003545/467292/Optical-Measurement-of-Thermal-Conductivity-Using},
	doi = {10.1115/1.4003545},
	number = {8},
	urldate = {2026-04-05},
	journal = {Journal of Heat Transfer},
	author = {Malen, Jonathan A. and Baheti, Kanhayalal and Tong, Tao and Zhao, Yang and Hudgings, Janice A. and Majumdar, Arun},
	month = aug,
	year = {2011},
	pages = {081601},
}

@article{wang_reducing_2022,
	title = {Reducing the uncertainty caused by the laser spot radius in frequency-domain thermoreflectance measurements of thermal properties},
	volume = {93},
	issn = {0034-6748, 1089-7623},
	url = {https://pubs.aip.org/rsi/article/93/2/023001/2843770/Reducing-the-uncertainty-caused-by-the-laser-spot},
	doi = {10.1063/5.0080119},
	number = {2},
	urldate = {2026-04-05},
	journal = {Review of Scientific Instruments},
	author = {Wang, Xiaoman and Jeong, Minyoung and McGaughey, Alan J. H. and Malen, Jonathan A.},
	month = feb,
	year = {2022},
	pages = {023001},
}

@article{collins_examining_2014,
	title = {Examining thermal transport through a frequency-domain representation of time-domain thermoreflectance data},
	volume = {85},
	issn = {0034-6748, 1089-7623},
	url = {https://pubs.aip.org/rsi/article/85/12/124903/109036/Examining-thermal-transport-through-a-frequency},
	doi = {10.1063/1.4903463},
	number = {12},
	urldate = {2026-04-05},
	journal = {Review of Scientific Instruments},
	author = {Collins, Kimberlee C. and Maznev, Alexei A. and Cuffe, John and Nelson, Keith A. and Chen, Gang},
	month = dec,
	year = {2014},
	pages = {124903},
}

@article{zhou_effective_2026,
	title = {Effective {Phonon} {Mean} {Free} {Paths} for {Thermal} {Modeling} of {Modern} {Semiconductors}},
	volume = {73},
	copyright = {https://ieeexplore.ieee.org/Xplorehelp/downloads/license-information/IEEE.html},
	issn = {0018-9383, 1557-9646},
	url = {https://ieeexplore.ieee.org/document/11410516/},
	doi = {10.1109/TED.2026.3665170},
	number = {4},
	urldate = {2026-04-05},
	journal = {IEEE Transactions on Electron Devices},
	author = {Zhou, Hao and Tiwari, Janak and Xu, Qihao and Islam, Md Fazla Rabbi Jim and Abedien, Tanvirul and Feng, Tianli},
	month = apr,
	year = {2026},
	pages = {2430--2439},
}

@article{rahman_simultaneous_2019,
	title = {Simultaneous measurement of anisotropic thermal conductivity and thermal boundary conductance of 2-dimensional materials},
	volume = {126},
	issn = {0021-8979, 1089-7550},
	url = {https://pubs.aip.org/jap/article/126/20/205103/595071/Simultaneous-measurement-of-anisotropic-thermal},
	doi = {10.1063/1.5118315},
	number = {20},
	urldate = {2026-09-20},
	journal = {Journal of Applied Physics},
	author = {Rahman, Mizanur and Shahzadeh, Mohammadreza and Pisana, Simone},
	month = nov,
	year = {2019},
	pages = {205103},
}

@article{tang_anisotropic_2021,
	title = {Anisotropic thermal conductivity tensor measurements using beam-offset frequency domain thermoreflectance ({BO}-{FDTR}) for materials lacking in-plane symmetry},
	volume = {164},
	issn = {00179310},
	url = {https://linkinghub.elsevier.com/retrieve/pii/S0017931020335365},
	doi = {10.1016/j.ijheatmasstransfer.2020.120600},
	urldate = {2026-09-20},
	journal = {International Journal of Heat and Mass Transfer},
	author = {Tang, Lei and Dames, Chris},
	month = jan,
	year = {2021},
	pages = {120600},
}

\appendix
\makeatletter
\makeatletter

\section{Simplifying 4-D integral in transfer function to 2-D integral}
\label{sec:appendix_4d_to_2d}
Eq.~(\ref{eq:transfer_function_4D}) is a four-dimensional integral, which can be simplified into a two-dimensional integral by applying the Hubbard–Stratonovich transformation,~\cite{schafer1980disordered}  which states that for a real, positive scalar $\gamma$,
\begin{equation}
    \label{eq:2D_identity}
    \int_{-\infty}^{\infty} \exp\left( -\frac{1}{2} \gamma x^2 + i\xi_x x\right) dx = \sqrt{\frac{2\pi}{\gamma}} \exp{\left(-\frac{\xi_x ^2}{2\gamma}\right)}.
\end{equation}
To convert Eq.~(\ref{eq:transfer_function_4D}) into the appropriate format, we use the identity $\cos(\theta) = \frac{\exp(i\theta)+\exp(-i\theta)}{2}$, and rewrite the cosine term in Eq.~(\ref{eq:transfer_function_4D}) to obtain 
\begin{multline}
    \label{eq:interm_1}
    H(\omega) =  \pi \delta(\omega-\omega_{\,0})\sum_{n=0}^{\infty} \sum_{m=0}^{\infty} \int_{-\infty}^{\infty} \int_{-\infty}^{\infty}\int_{-\infty}^{\infty} \int_{-\infty}^{\infty} \frac{a_ma_n}{\pi r_1^2}\left( -\frac{D}{C}\right) \\ \times \left\{\exp\left(\frac{-(\xi_x-\frac{2n\pi}{L_x})^2 r_0^2}{8}\right) + \exp\left(\frac{-(\xi_x + \frac{2n\pi}{L_x})^2 r_0^2}{8}\right)\right\} \\ \times \exp \left( i\xi_x x + i \xi_y y -\frac{ \xi_y^2 r_0^2}{8} - \frac{2(x^2+y^2)}{r_1^2}\right) \\ \times \frac{1}{2}\left\{\exp \left( i\frac{2m\pi x}{L_x}\right) + \exp \left( -i\frac{2m\pi x}{L_x}\right)\right\} d\xi_x d\xi_y dx dy.
\end{multline}
Noting that $\xi_y$ is independent of y, the integration of the variable $y$ can be rearranged in the form of Eq.~(\ref{eq:2D_identity}) as
\begin{multline}
    \label{eq:interm_2}
     H(\omega) =  \pi \delta(\omega-\omega_{\,0})\sum_{n=0}^{\infty} \sum_{m=0}^{\infty}  \int_{-\infty}^{\infty}\int_{-\infty}^{\infty} \int_{-\infty}^{\infty} \frac{a_ma_n}{2\pi r_1^2}\left( -\frac{D}{C}\right) \\ \times \left\{\exp\left(\frac{-(\xi_x-\frac{2n\pi}{L_x})^2 r_0^2}{8}\right) + \exp\left(\frac{-(\xi_x + \frac{2n\pi}{L_x})^2 r_0^2}{8}\right)\right\}  \times \exp \left( i\xi_x x  - \frac{2x^2}{r_1^2}\right) \\ \times \left\{\exp \left( i\frac{2m\pi x}{L_x}\right) + \exp \left( -i\frac{2m\pi x}{L_x}\right)\right\} \left[\exp \left(-\frac{ \xi_y^2 r_0^2}{8}\right)\int_{-\infty}^{\infty} \exp \left( i \xi_y y  - \frac{2y^2}{r_1^2}\right)dy\right]d\xi_x d\xi_y dx.
\end{multline}
With $\gamma = \frac{4}{r_1^2}$, the expression simplifies to
\begin{multline}
    \label{eq:interm_3}
     H(\omega) =  \pi \delta(\omega-\omega_{\,0})\sum_{n=0}^{\infty} \sum_{m=0}^{\infty}  \int_{-\infty}^{\infty}\int_{-\infty}^{\infty} \int_{-\infty}^{\infty} \frac{a_ma_n}{2\pi r_1^2}\left( -\frac{D}{C}\right) \\ \times \left\{\exp\left(\frac{-(\xi_x-\frac{2n\pi}{L_x})^2 r_0^2}{8}\right) + \exp\left(\frac{-(\xi_x + \frac{2n\pi}{L_x})^2 r_0^2}{8}\right)\right\} \times \exp \left( i\xi_x x  - \frac{2x^2}{r_1^2}\right) \\ \times \left\{\exp \left( i\frac{2m\pi x}{L_x}\right) + \exp \left( -i\frac{2m\pi x}{L_x}\right)\right\} \left[\exp \left(-\frac{ \xi_y^2 r_0^2}{8}\right) \sqrt{\frac{\pi r^2}{2}} \exp \left(-\frac{\xi_y^2 r_1^2}{8}\right)\right]d\xi_x d\xi_y dx.
\end{multline}
Further simplification and rearranging the terms leads to
\begin{multline}
    \label{eq:interm_4}
     H(\omega) =  \pi \delta(\omega-\omega_{\,0})\sum_{n=0}^{\infty} \sum_{m=0}^{\infty}  \int_{-\infty}^{\infty}\int_{-\infty}^{\infty} \int_{-\infty}^{\infty} \frac{a_ma_n}{2\sqrt{2\pi r_1^2}}\exp \left(-\frac{ \xi_y^2}{8} (r_0^2 + r_1^2)\right)\left( -\frac{D}{C}\right) \\ \times \left\{\exp\left(\frac{-(\xi_x-\frac{2n\pi}{L_x})^2 r_0^2}{8}\right) + \exp\left(\frac{-(\xi_x + \frac{2n\pi}{L_x})^2 r_0^2}{8}\right)\right\} \times \exp \left( i\xi_x x  - \frac{2x^2}{r_1^2}\right) \\ \times \left\{\exp \left( i\frac{2m\pi x}{L_x}\right) + \exp \left( -i\frac{2m\pi x}{L_x}\right)\right\} d\xi_x d\xi_y dx.
\end{multline}
Similarly, integration over $x$ can be performed by collecting terms as 
\begin{multline}
    \label{eq:interm_5}
     H(\omega) =  \pi \delta(\omega-\omega_{\,0})\sum_{n=0}^{\infty} \sum_{m=0}^{\infty}  \int_{-\infty}^{\infty} \int_{-\infty}^{\infty} \frac{a_ma_n}{2\sqrt{2\pi r_1^2}} \exp\left(-\frac{ \xi_y^2}{8} (r_0^2 + r_1^2)\right)\left( -\frac{D}{C}\right) \\ \times \left\{\exp\left(\frac{-(\xi_x-\frac{2n\pi}{L_x})^2 r_0^2}{8}\right) + \exp\left(\frac{-(\xi_x + \frac{2n\pi}{L_x})^2 r_0^2}{8}\right)\right\} \\ \times \left[\int_{-\infty}^{\infty} \exp \left\{ i\left(\xi_x + \frac{2m\pi}{L_x}  \right)x  - \frac{2x^2}{r_1^2}\right\} dx + \int_{-\infty}^{\infty} \exp \left\{ i\left(\xi_x - \frac{2m\pi}{L_x}  \right)x  - \frac{2x^2}{r_1^2}\right\} dx\right]d\xi_x d\xi_y.
\end{multline}
Again, using the identity in Eq.~(\ref{eq:2D_identity}) to perform integration on the $x$ coordinates leads to the expression of $H(\omega)$ in a two-dimensional integral form as
\begin{multline}
    \label{eq:2D_transfer_function_unequal}
    H(\omega) = \pi \delta(\omega-\omega_{\,0})\sum_{n=0}^{\infty} \sum_{m=0}^{\infty} \int_{-\infty}^{\infty} \int_{-\infty}^{\infty} \frac{a_na_m}{4} \left(-\frac{D}{C}\right) \exp \left\{-\frac{ \xi_y^2}{8} (r_0^2 + r_1^2)\right\} \\ \times \left[\exp\left\{ -\frac{r_0^2}{8} \left(\xi_x-\frac{2n\pi}{L_x}\right)^2 -\frac{r_1^2}{8} \left(\xi_x-\frac{2m\pi}{L_x}\right)^2\right\} \right. \\ \left. + \exp\left\{ -\frac{r_0^2}{8} \left(\xi_x-\frac{2n\pi}{L_x}\right)^2 -\frac{r_1^2}{8} \left(\xi_x+\frac{2m\pi}{L_x}\right)^2\right\} \right. \\ \left. + \exp\left\{ -\frac{r_0^2}{8} \left(\xi_x+\frac{2n\pi}{L_x}\right)^2 -\frac{r_1^2}{8} \left(\xi_x-\frac{2m\pi}{L_x}\right)^2\right\} \right. \\ \left. + \exp\left\{ -\frac{r_0^2}{8} \left(\xi_x+\frac{2n\pi}{L_x}\right)^2 -\frac{r_1^2}{8} \left(\xi_x+\frac{2m\pi}{L_x}\right)^2\right\}\right]  d\xi_x d\xi_y.
\end{multline}

\section{Variation of predicted phase with laser spot radius}
\label{sec:appendix_spot_size_effect}
\begin{figure}[h!]
 \centering
 \includegraphics[width= 0.8\linewidth]{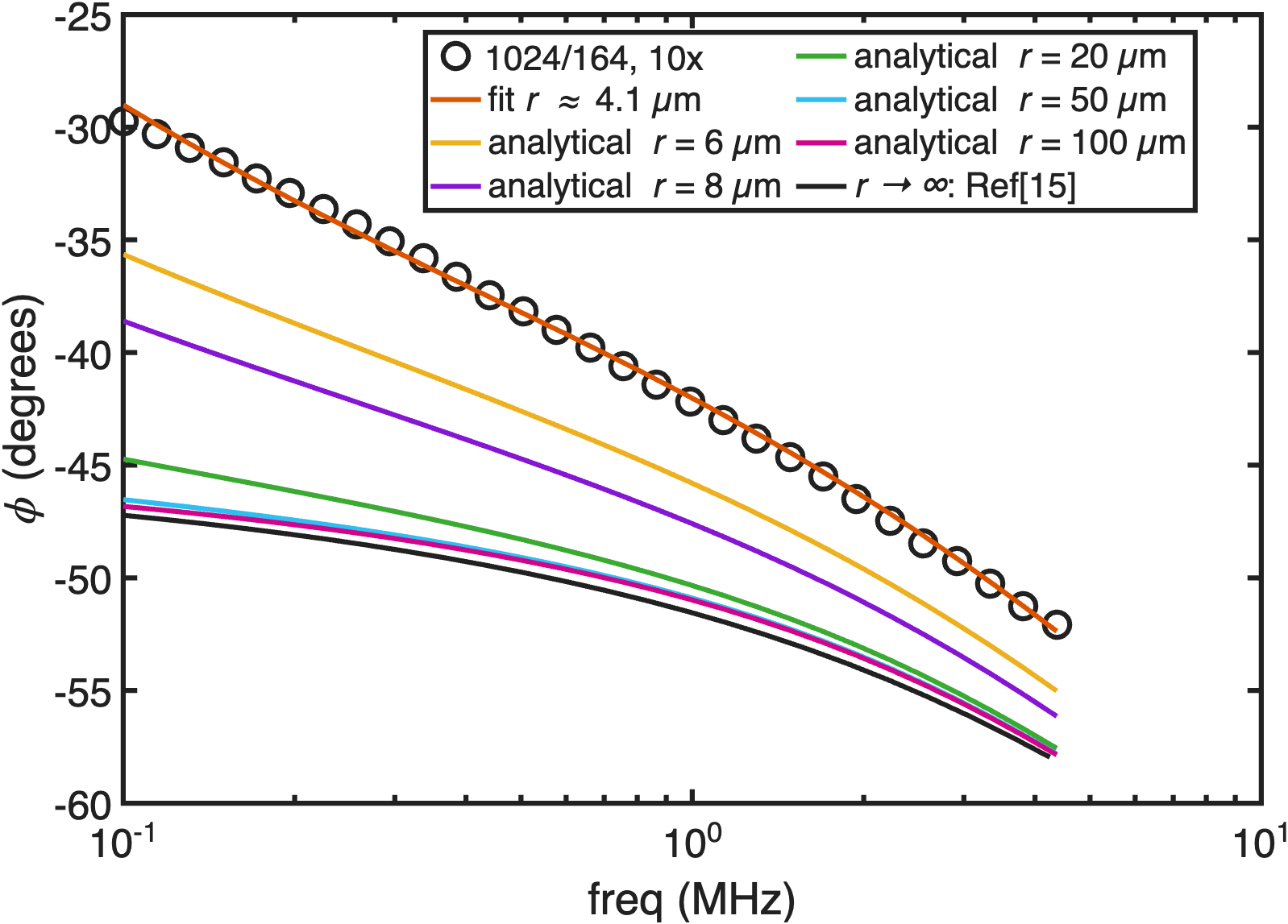}
 \caption{Spot radius dependence of predicted phase on nanolines ($w$ = 1024 nm and gap $(L-w)$ = 164 nm) deposited on a-SiO$_2$ sample from analytical model derived in this work compared against previously used model ($r \rightarrow \infty$) in TDTR study.~\cite{hu2015spectral}}
 \label{fig:spot_size_dependence}
\end{figure}

In Fig.~\ref{fig:spot_size_dependence}, we show how the predicted phase changes upon varying the spot radius for a nanoline pattern on the a-SiO$_2$ substrate with $w = 1024$ nm and $L=1188$ nm.
As we increase the spot radius in our model while keeping other parameters constant, the predicted phase lag increases for all frequencies. Eventually, for a large radius ($r>100\,\mu$m), the predicted phase converges to the values predicted by the previously used infinite radius model that was adopted for TDTR measurements.~\cite{hu2015spectral}
Our analytical model, used with the bulk thermal conductivity of a-SiO$_2$, shows excellent agreement with the measurements, further reinforcing the impact of the finite spot radius.

\section*{Supplementary Material}
\renewcommand{\appendixname}{}
\renewcommand{\theequation}{S\arabic{equation}}
\setcounter{equation}{0} 
\renewcommand{\thefigure}{S\arabic{figure}}
 \setcounter{figure}{0}
\renewcommand{\thetable}{S\arabic{table}}
\setcounter{table}{0}
\renewcommand{\thesection}{S\arabic{section}}
\setcounter{section}{0} 
{\color{black}
\section{Fabrication of nanoline array}\label{SI:nanofab}

\begin{figure}[h] \label{SI:fabprocess}
    \centering
    \includegraphics[width=0.9\linewidth]{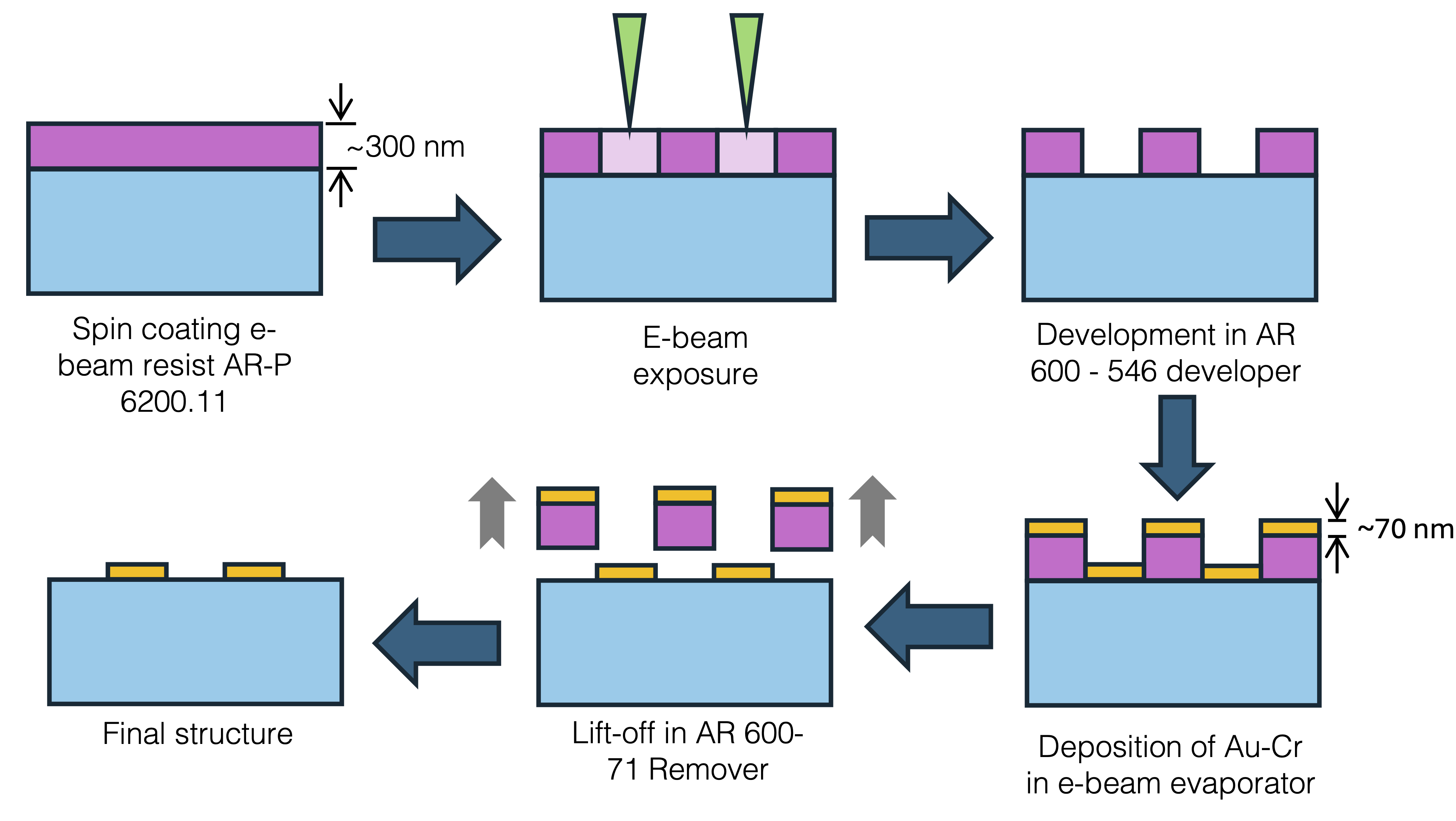}
    \caption{\textbf{(a)} Nano-fabrication process for metallic heaters.}
    \label{fig:SI-nanofab}
\end{figure}

To test the validity of the model developed in Sec.~\ref{sec:analytical}, we deposited Au nanolines on $c$-sapphire and a-SiO$_2$ substrates and measured their responses using FDTR.
The electron-beam lithography process followed by a lift-off procedure that was used to fabricate the nanoline patterns is illustrated in Fig.~\ref{SI:fabprocess}.
The targeted nanoline widths range from $100$ to $1000$ nm, with a $200$ nm gap between adjacent lines.
In the direction parallel to the lines [$y$ in Fig.~\ref{fig:description}(b)], the length of the nanolines was set to 50 $\mu$m with a periodicity of 100 $\mu$m, as shown in Fig.~\ref{fig:nanofab}(a). 
For every value of nanoline width, a 500 $\mu$m $\times$ 450 $\mu$m area was patterned. 

A double-side polished 400 $\mu$m thick $2$-inch wafer of $c$-sapphire, purchased from  MSE Supplies LLC, was cleaned by sonication in acetone for 10 minutes, followed by rinsing with isopropyl alcohol (IPA) and drying with nitrogen gas.  
To promote the adhesion of e-beam resist, the substrate was baked at 180$^\circ$C for $30$ minutes, followed by an O$_2$ plasma cleaning step in an IPC barrel etcher operated at $750$ mTorr and $300$ W for $10$ minutes.
The baking process removes surface-adsorbed moisture, whereas O$_2$ plasma eliminates residual organic contaminants.
An approximately $300$ nm layer of e-beam resist (AR-P 6200.11, CSAR) was spin-coated onto the substrate, followed by a pre-bake at 150$^\circ$C for $1$ minute. 
Subsequently, an approximately $20$ nm conductive film of DisCharge-H$_2$O-2X (Dischem Inc.) was spin-coated to dissipate surface charge during electron-beam exposure and to prevent pattern distortion. 
The lithography was performed using a G100 Electron Beam Lithography system operating at an accelerating voltage of $100$ kV with a spot radius of $1.8$ nm.
Based on the manufacturer-recommended process, the dose-to-clear for CSAR was determined to be 120 $\mu$C/cm$^2$ for a $300$ nm thick layer on a $c$-sapphire substrate.
We used a higher dose of 200 $\mu$C/cm$^2$ to account for proximity effects and to achieve a slightly negative resist-sidewall profile after development, which facilitates an efficient lift-off process.
The same parameters were found to be suitable for the 300 $\mu$m thick $2$-inch a-SiO$_2$ substrate.

Following exposure, the DisCharge-H$_2$O-2X coating was removed by rinsing with deionized water and the substrate was dried under a nitrogen stream. 
Pattern development was performed at room temperature using AR600-546 (Allresist) developer for $2$ minutes, followed by immersion in IPA for 1 minute to stop the reaction. 
The patterned substrate was then treated with a mild dose of O$_2$ plasma ($20$ W, $20$ mTorr) in a Plasma-therm 790 reactive ion etching (RIE) system for $40$ seconds to remove residual resist and enhance metal adhesion in the subsequent step.
A $5$ nm thick Cr adhesion layer~\cite{jeong_enhancement_2016} followed by a $70$ nm Au layer were deposited on the patterns in a Kurt J. Lesker PRO Line PVD75 E-beam Evaporator under high vacuum ($\approx 3 \times 10^{-7}$ Pa).
The lift-off process was performed by upside-down immersion of the samples overnight in AR600-71 Remover. 
Although a spatially uniform increase in dose preserved the shape-integrity of the patterns, an artifact of this over-exposure is that the nanolines are about $20$ nm too wide on a-SiO$_2$ and $35$-$50$ nm too wide on $c$-sapphire.
}

\section{Determination of \NoCaseChange{Au-Cr} thickness}
\label{sec:SI_profilometer}
\begin{figure}[h]
    \centering
    \includegraphics[width=0.8\linewidth]{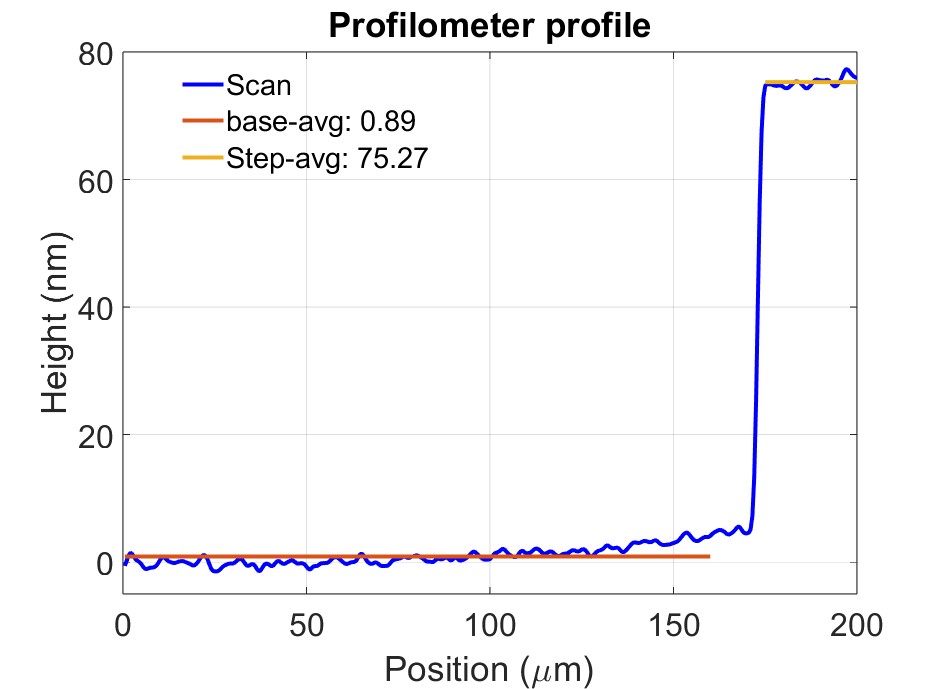}
    \caption{Sample profilometer scan.}
    \label{fig:SI_profilometer}
\end{figure}
Fig.~\ref{fig:SI_profilometer} shows a sample profilometer scan measured with KLA Tencor P-15 Profilometer on a $c$-sapphire square patch to determine the combined thickness of Au and Cr layers. The combined thickness was determined by subtracting the base average height from the step average height, which was measured to be $\approx$ 75 nm.
\FloatBarrier
{\color{black}
\section{Finite element validation of heat flux boundary condition on continuous film}\label{sec:nanoline_FEM}

\begin{figure}[h!]
    \centering
    \includegraphics[width=0.8\linewidth]{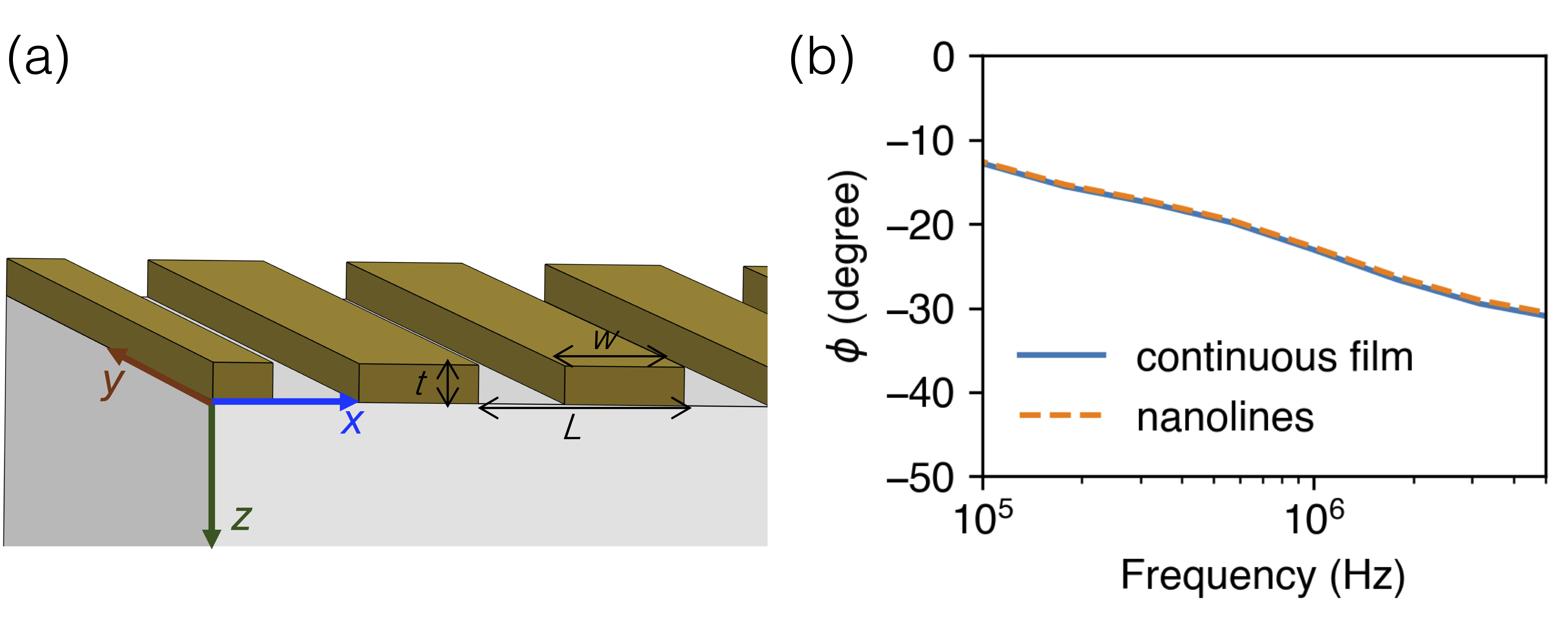}
    \caption{Comparison between the phase lag recorded in the FEM model using continuous film simplification vs physically modeling metal nanolines \textbf{(a)} Schematic diagram of the FEM model geometry with $w=234$ nm, $L=403$ nm, $t=75$ nm. \textbf{(b)} The phase lag shows excellent match between the two models.}
    \label{fig:SI_nanoline_simplification}
\end{figure}

The analytical model derived in Sec.~\ref{sec:analytical} approximates the nanoline geometry heated with a Gaussian laser spot by applying a modified boundary heat flux profile on a continuous Au film.
The presence of a physical gap is simulated by setting $k_x = 0$ for the Au film.
To evaluate the effect of this simplifying assumption on the predicted phase lag, we prepared a 3-D finite element model of nanoline on the substrate, as shown in Fig.~\ref{fig:SI_nanoline_simplification}(a).
Particularly, we evaluate whether the backflow of heat from the substrate to the continuous Au film and its spread in the $y$-direction in the Au film affect the phase lag.

We used COMSOL to perform a steady-state solution under constant, spatially Gaussian heat flux followed by a harmonic perturbation analysis at several modulation frequencies encompassing the whole range used in our measurements.
A 3-D finite element model is computationally expensive, and a few simplifications are made to keep the simulation computationally tractable.
The size of the computational domain is limited to $L_x=8\,\mu$m in the $ x$-direction and $L_y=5\,\mu$m in the $y$-direction, and the substrate is only modeled as a $d=5\,\mu$m thick layer in the $ z$-direction.
Symmetry boundary conditions are applied for the $Y$-$Z$ plane at $x=0$ and the $X$-$Z$ plane at $y=0$, such that only a quarter model is solved.
To achieve a steady-state solution quickly, isothermal boundary conditions are applied at $x=L_x$ and $z=d$.
However, $y=L_y$ boundary is kept insulated to avoid forced heat flow in the $y$-direction.
These simplifications are chosen to speed up the finite element simulation without affecting the core conclusions regarding the backflow of the heat from the substrate and the $y$-direction heat flow in the Au film.

We performed the same analysis by replacing the physical nanolines with a continuous film and applied the same modified heat flux as used in deriving the analytical model (Sec. \ref{sec:analytical}) and setting $k_x=0$ in the film.
Both the simulations used identical material properties except $k_x$ of the Au film.
Fig.~\ref{fig:SI_nanoline_simplification}(b) shows the comparison between the phase lag computed for the two models.
The maximum absolute deviation in the phase lag between the two simulations is less than $1.1^{\circ}$ over the entire frequency range. 
The phase shift corresponds to a 0.3\% and 2.9\% change in the extracted thermal conductivity of c-sapphire and the thermal boundary conductance at the Au-Cr/c-sapphire interface, respectively. 
The analysis clearly shows that the assumption of a continuous Au film with $k_x=0$ excellently approximates the nanolines with a physical gap for our measurements.
}

\section{Absence of thermoreflectance from bare substrate}\label{SI:absence}
\begin{figure}[h!]
    \centering
    \includegraphics[width=0.8\linewidth]{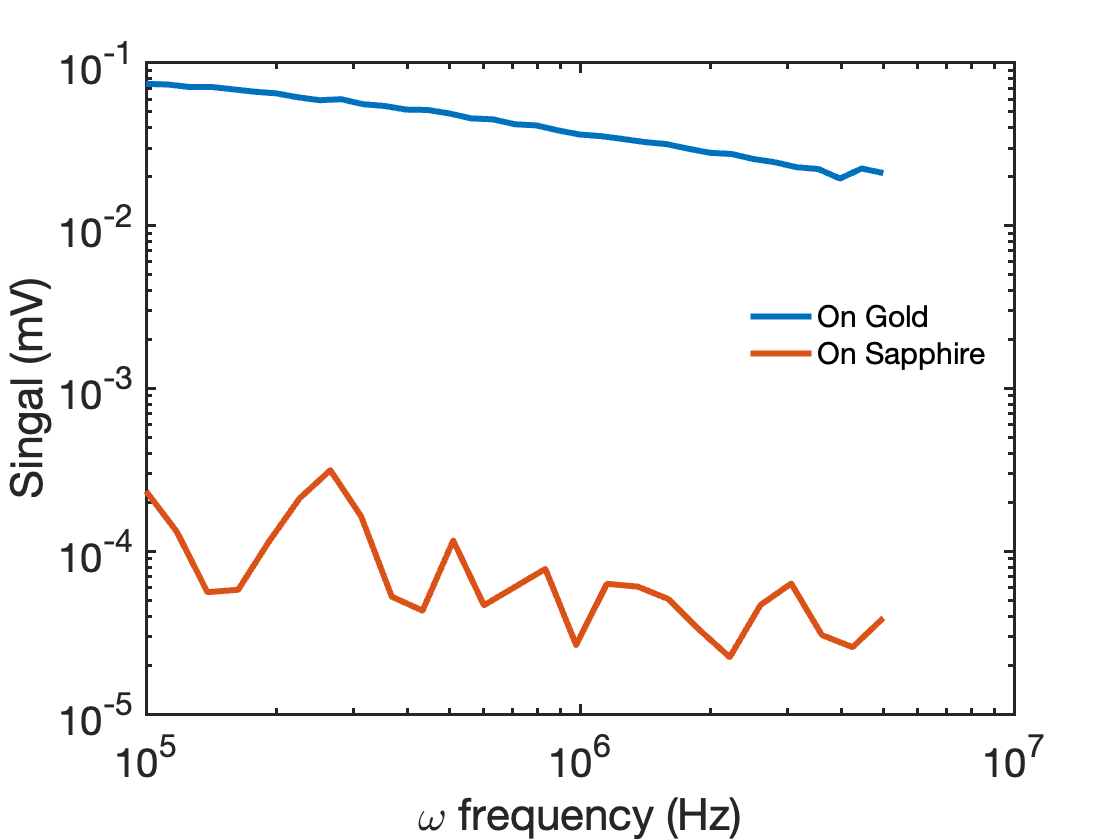}
    \caption{Thermoreflectance signal recorded on $c$-sapphire sample on the area covered with an Au transducer and on the bare $c$-sapphire substrate.}
    \label{fig:SI_thermoreflectance_sapphire}
\end{figure}
Fig.~\ref{fig:SI_thermoreflectance_sapphire} shows the thermoreflectance signal obtained on the Au transducer deposited on the $c$-sapphire substrate and on the bare $c$-sapphire substrate. Since $c$-sapphire is transparent to the pump (405 nm)  and probe (532 nm) lasers, it does not absorb heat, and the signal obtained for the bare substrate is basically coming from the instrumentation noise of the setup. This also ensures that the laser passing through the gap between the nanolines and incident directly onto the substrate does not affect the thermoreflectance measurement.

\section{Phase lag and analytical solution fitting of square patches}
\begin{figure}[h!]
    \centering
    \includegraphics[width=\linewidth]{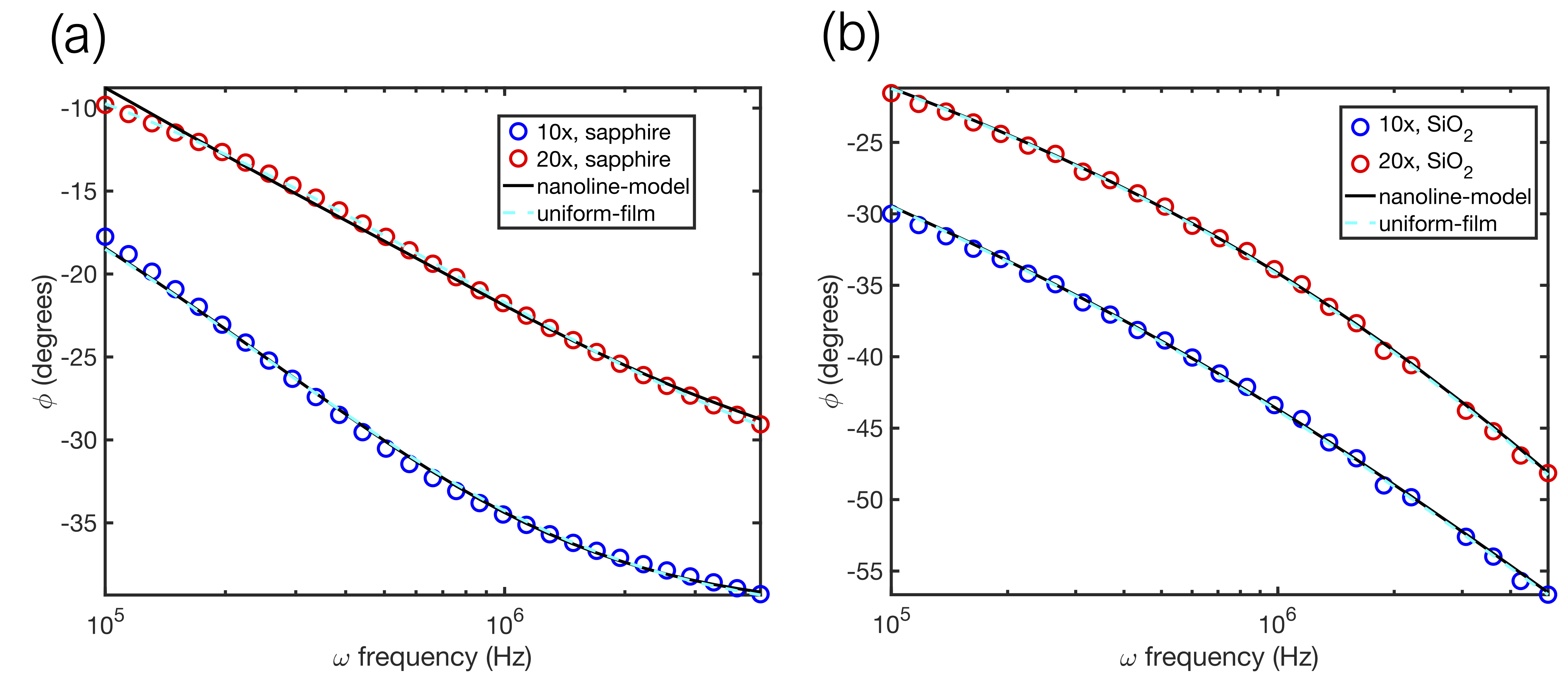}
    \caption{Phase lag and analytical solution fitting of square patch on \textbf{(a)} $c$-sapphire substrate and \textbf{(b)} a-SiO$_2$ substrate.}
    \label{fig:SI_square_sapphire}
\end{figure}
Figs.~\ref{fig:SI_square_sapphire}(a) and (b) show the fitting of the phase lag obtained in the FDTR measurement on one of the square patches deposited on $c$-sapphire and a-SiO$_2$, respectively.
As the size of the square patches (50 $\mu m \times$ 50 $\mu m$) is large compared to the spot radius used in the measurements, the conventional FDTR model with an infinite uniform transducer film is used for fitting the measurement. For comparison, we also fit the measurements using our analytical solution for the periodic nanoline patterns with $w=50\,\mu$m and $L=100\,\mu$m. The excellent agreement between the two curves shows that our analytical model converges to the standard FDTR solution in the limit of large $w$ relative to the spot radius. 

\renewcommand{\arraystretch}{1.3}
\begin{table}[htbp]
    \caption{a-SiO$_2$ square patch fitting with nanoline model and uniform film model.}
    \begin{center}
        \begin{tabular}{|c|c|c|c|c|}
        \hline
        \multirow{2}{*}{\centering \textbf{model}}& \multicolumn{2}{c|}{\textbf{\textit{$k_{y,Au}$ [W/m-K]}}} & \multicolumn{2}{c|}{\textbf{\textit{$TBC$ [MW/m$^2$K]}}} \\ \cline{2-5}
            & \textbf{10x} & \textbf{20x} & \textbf{10x} & \textbf{20x}\\
            \hline
            uniform&$160 \pm 4.8$$^{\mathrm{a}}$ & $160 \pm 4.8$$^{\mathrm{a}}$ & $42^{+6.8}_{-14}$ & $32^{+12}_{-5.3}$\\
            nanoline&$160 \pm 4.8$$^{\mathrm{a}}$ & $160 \pm 4.8$$^{\mathrm{a}}$ & $39^{+6.3}_{-4.1}$ & $37^{7.1}_{3.2}$\\   
            \hline
            \multicolumn{4}{l}{$^{\mathrm{a}}$ Uncertainty from 4PP method}
        \end{tabular}
        \label{square_to_nanoline_fused_silica}
    \end{center}
\end{table}

\renewcommand{\arraystretch}{1.3}
\begin{table}[htbp]
    \caption{$c$-sapphire square patch fitting with nanoline model and uniform film model.}
    \begin{center}
        \begin{tabular}{|c|c|c|c|c|}
        \hline
        \multirow{2}{*}{\centering \textbf{model}}& \multicolumn{2}{c|}{\textbf{\textit{$k_{c-sapphire}$ [W/m-K]}}} & \multicolumn{2}{c|}{\textbf{\textit{$TBC$ [MW/m$^2$K]}}} \\ \cline{2-5}
            & \textbf{10x} & \textbf{20x} & \textbf{10x} & \textbf{20x}\\
            \hline
            uniform & $37^{+2.8}_{-2.6}$ & $37^{+2.7}_{-2.7}$  & $338^{+20}_{-19}$ & $429^{+20}_{-17}$\\
            nanoline&$38^{+2.5}_{-2.3}$ & $38^{+2.6}_{-2.6}$  & $286^{+22}_{-17}$& $392^{+28}_{-33}$ \\   
            \hline
        \end{tabular}
        \label{square_to_nanoline_sapphire}
    \end{center}
\end{table}

\section{Transmitted energy simulation}\label{SI:transmit}

\begin{figure}[h]
    \centering
    \includegraphics[width=0.9\linewidth]{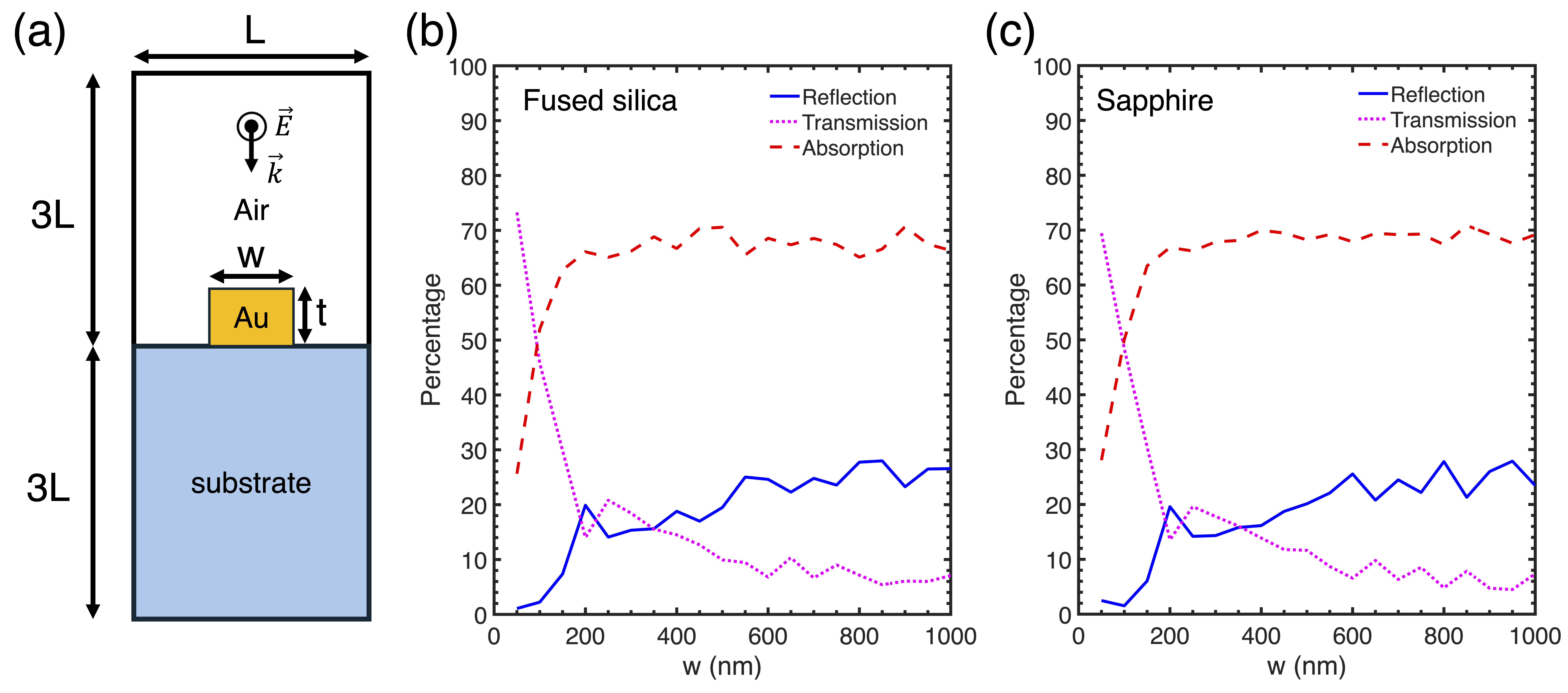}
    \caption{\textbf{(a)} Schematics of the simulation domain. The periodicity $L$ is chosen such that the gap between adjacent nanolines is $200$ nm. \textbf{(b)} Calculated percentage of reflected, transmitted, and absorbed laser energy by Au gratings on a-SiO$_2$ substrate and \textbf{(c)} $c$-sapphire substrate.}
    \label{fig:comsol}
\end{figure}

 Fig.~\ref{fig:comsol}(a) shows the schematic of the simulation domain used to calculate the transmitted laser energy through the gap between the adjacent metal nanolines using the COMSOL wave optics module.~\cite{COMSOL} One unit cell of the periodic grating structure was simulated with Floquet boundary conditions applied on the lateral boundaries. The laser is incident perpendicular to the grating surface from air, $\vec{k}$ shows the wavevector of the incident electromagnetic wave. The polarization electric field $\vec{E}$ is aligned with the longitudinal direction (into the page in Fig.~\ref {fig:comsol}). The top boundary is used for applying wave excitation due to a unit power incident electromagnetic beam. Top and bottom boundaries are used to measure the reflected and transmitted power, respectively.
 
 The optical properties of Au nanolines, air and substrates were obtained from `Refractiveindex. info database of optical constants'~\cite{polyanskiy2024refractiveindex} for the pump laser wavelength (405 nm), and are listed in Table~\ref{table:comsol_mat}. To be consistent with our nanoline pattern geometry, the gap between nanolines was kept at 200 nm, and the Au thickness was kept at 75 nm. Fig.~\ref{fig:comsol}(b) and (c) show how the percentage of incident light reflected, transmitted, and absorbed by the Au nanoline grating changes with nanoline width. The results obtained for a-SiO$_2$ and $c$-sapphire are qualitatively similar. For a smaller nanoline width, the length of the gap forms a large fraction of the unit cell, and significant light is transmitted to the substrate. As the nanoline width relative to the periodicity increases, the absorption in the Au nanoline rapidly increases and saturates to $\approx 60\%$. For larger nanolines, the decreasing trend of transmission power and the increasing trend of reflected power balance each other.

 Although there is $\approx 20\%$ transmission for the smallest nanoline width ($200$ nm) used in our study, both a-SiO$_2$ and $c$-sapphire are transparent to the pump wavelength, and no direct substrate heating is expected. Indirect heating of Au nanolines from the light reflected back from the bottom surface is possible. The application of a refractive index-matching liquid can be used to minimize reflection from the bottom surface. We used a thin layer of glycerine (refractive index $\approx 1.5$) to achieve this effect.

\renewcommand{\arraystretch}{1.3}
\begin{table}[htbp]
    \caption{Material properties for COMSOL simulation for 405 nm pump laser.~\cite{polyanskiy2024refractiveindex}}
    \begin{center}
        \begin{tabular}{|c|c|c|c|c|}
        \hline
        Material & Real part of refractive index & Imaginary part of refractive index \\
            \hline
            Au & 1.4652 & 1.9549 \\
            a-SiO$_2$ & 1.4696 & 0 \\  
            $c$-sapphire & 1.7857 & 0 \\ 
            \hline
        \end{tabular}
        \label{table:comsol_mat}
    \end{center}
\end{table}

{\color{black}
\section{Uncertainty Quantification of Laser Spot Positioning}\label{SI:laserspot}
\begin{figure}[!h]
    \centering
    \includegraphics[width=1\linewidth]{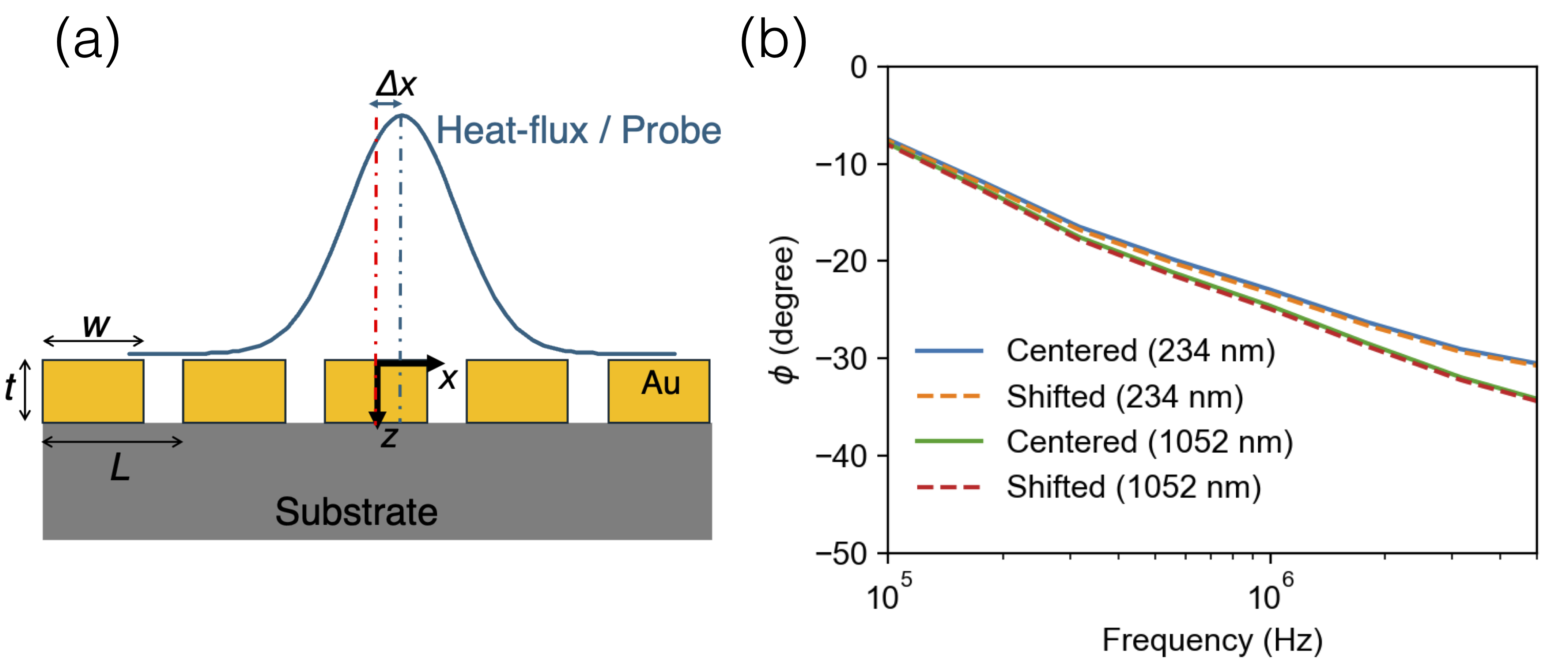}
    \caption{\textbf{(a)} Schematics of the offset between the nanoline centerline and the center of the laser profile. \textbf{(b)} FEM-calculated phase responses of the 234 nm and 1052 nm nanoline patterns on sapphire with and without a 100 nm lateral offset from the nanoline centerline.}
    \label{fig:offcenter}
\end{figure}

To position the laser spot on the nanolines, we used a micro-manipulator to translate the sample relative to the fixed laser spot while monitoring the measured signal. The laser spot was aligned to the nanoline centerline by maximizing the signal amplitude. Because the micro-manipulator has a positioning resolution of 50 nm, a small residual lateral misalignment between the laser spot and the nanoline centerline may remain. We therefore evaluated the influence of this uncertainty on the measured phase using Finite Element Method (FEM) simulations.

In the simulations, the laser spot was displaced from the nanoline centerline along the x-direction, and the resulting phase at the measurement frequencies was calculated. We considered a conservative worst-case lateral offset of 100 nm, corresponding to twice the micro-manipulator resolution. For both the 234 nm and 1052 nm sapphire nanoline patterns, the calculated phase variation was less than $0.36^{\circ}$ over the entire frequency range. The phase shift corresponds to a 2.1\% and 0.7\% change in the extracted thermal conductivity of $c$-sapphire and the thermal boundary conductance at the Au-Cr/$c$-sapphire interface, respectively.}

\end{document}